\documentclass[a4paper]{article} % must use LaTeX2e

\usepackage{amssymb}
\usepackage{amsmath}
\usepackage{graphicx}
\usepackage{subfigure}
\usepackage{psfrag}

\newcommand\comment[1]{} % invisible remark in manuscript
\newcommand{\sizedef}{
      \headheight=0pt                               % zero space header
	  \topmargin=-1.5cm \headsep=1.5cm              % put near top of page
      \oddsidemargin=-0.5cm \evensidemargin=-0.5cm  % adjust left
      \textheight=22truecm \textwidth=16.5truecm    % PLAIN text dimensions
}
\newtoks\reportnoregister \newtoks\eprintnoregister
\newcommand{\reportnumber}[1]{\reportnoregister={#1}}
\newcommand{\eprintnumber}[1]{\eprintnoregister={#1}}

\reportnumber{\mbox{}} % empty registers to start with, needed if the
\eprintnumber{\mbox{}} % \reportnumber (or \eprintnumber) command isn't used

\newcommand{\reportid}{
   \begin{minipage}{17cm}\vspace{-3.2cm}
     \begin{flushright}
      {\normalsize \the\reportnoregister \\[-.2cm]
	    \eprint{\the\eprintnoregister}}\vspace{3.2cm}
     \end{flushright}
   \end{minipage}\hspace{-17cm} }

\catcode`@=11   % make @ act like letter temporarily
\def\title#1{\gdef\@title{\reportid#1}}
\catcode`@=12   % back to nonletter category

\newcommand{\eprint}{\textsf} % style for e-print references
\newcommand{\eref}[1]{E-print: \hbox{\eprint #1}} % for e-print references

\newcommand{\journalfont}{\rm}  % this allows redefinition of the font later
\newcommand{\jou}[1]{{\journalfont #1\ }}
\newcommand{\joudef}[2]{\newcommand #1{\jou{\ignorespaces #2}}}

\joudef{\aaa}    { Astron.\ Astrophys.}
\joudef{\aip}    { Adv.\ Phys.}
\joudef{\adm}    { Adv.\ Math.}
\joudef{\am}     { Ann.\ Math.}
\joudef{\apny}   { Ann.\ Phys.\ (N.Y.)}
\joudef{\apj}    { Astrophys.\ J.}
\joudef{\apjs}   { Astrophys.\ J.\ Suppl.}
\joudef{\cjp}    { Can.\ J.\ Phys.}
\joudef{\cmp}    { Commun.\ Math.\ Phys.}
\joudef{\cqg}    { Class.\ Quantum Grav.}
\joudef{\faa}    { Funct.\ Anal.\ Appl.}
\joudef{\grg}    { Gen.\ Rel.\ Grav.}
\joudef{\ijmpd}  { Int.\ J.\ Mod.\ Phys.\ D}
\joudef{\ijtp}   { Int.\ J.\ Theor.\ Phys.}
\joudef{\invm}   { Invent.\ Math.}
\joudef{\jm}     { J.\ Math.}
\joudef{\jmp}    { J.\ Math.\ Phys.}
\joudef{\jpa}    { J.\ Phys.\ A}
\joudef{\mnras}  { Mon.\ Not.\ R.\ Ast.\ Soc.}
\joudef{\mpla}   { Mod.\ Phys.\ Lett.\ A} 
\joudef{\nature} { Nature}
\joudef{\nc}     { Nuovo Cim.}
\joudef{\npb}    { Nuc.\ Phys.\ B}
\joudef{\ph}     { Physica}
\joudef{\pla}    { Phys.\ Lett. A}
\joudef{\plb}    { Phys.\ Lett. B}
\joudef{\pr}     { Phys.\ Rev.}
\joudef{\prd}    { Phys.\ Rev.\ D}
\joudef{\prep}   { Phys.\ Rep.}
\joudef{\prl}    { Phys.\ Rev.\ Lett.}
\joudef{\prsla}  { Proc.\ Roy.\ Soc.\ Lond.\ A}
\joudef{\ptp}    { Prog.\ Theor.\ Phys.}
\joudef{\ptps}   { Prog.\ Theor.\ Phys.\ Suppl.}
\joudef\rmp      { Rev.\ Mod.\ Phys.}
\joudef\spj      { Sov.\ Phys.\ JETP}
\catcode`@=11

\newcommand\eqalign[1]{\null\,\vcenter{\openup\jot\m@th
  \ialign{\strut\hfil$\displaystyle{##}$&$\displaystyle{{}##}$\hfil
      \crcr#1\crcr}}\,}
\newcommand\meqalign[1]{\null\,\vcenter{\openup\jot\m@th
  \ialign{\strut\hfil$\displaystyle{##}$&&$\displaystyle{{}##}$\hfil
      \crcr#1\crcr}}\,}
\def\ps@reportnumber{%
    \let\@oddfoot\@empty\let\@evenfoot\@empty
    \def\@oddhead{\hfil\rightmark}}
	
\catcode`@=12   % back to nonletter category
\newdimen\arrayruleHwidth
\makeatletter
\newcommand\Hline{\noalign{\ifnum0=`}\fi\hrule \@height \arrayruleHwidth
  \futurelet \@tempa\@xhline}
\makeatother
\newcommand\thickbaselines{\baselineskip=20pt\lineskip=3pt\lineskiplimit=3pt}

\catcode`@=11

\renewcommand\matrix[1]{\null\,\vcenter{\thickbaselines\m@th
    \ialign{\hfil$##$\hfil&&\quad\hfil$##$\hfil\crcr
      \mathstrut\crcr\noalign{\kern-\baselineskip}
      #1\crcr\mathstrut\crcr\noalign{\kern-\baselineskip}}}\,} 
\catcode`@=12   % back to nonletter category

\newcommand\be{\begin{equation}} \newcommand\ee{\end{equation}} % numbered
\newcommand\bd{\begin{displaymath}}\newcommand\ed{\end{displaymath}}%unnumb'd

\renewcommand{\d}{{\rm d}} % Note: \d is built-in TeX command for underdot
\newcommand{\e}{{\rm e}}

\newcommand\Oscr{{\cal O}}

\newcommand\undersim[1]{\mathop{\vtop{\ialign{##\crcr
     $\hfil\displaystyle{#1}\hfil$\crcr\noalign
     {\kern1pt\nointerlineskip}\hbox{$\hfil\sim\hfil$}\crcr
     \noalign{\kern1pt}}}}}

\newcommand\case[2]{\textstyle{\frac{#1}{#2}}}
\newcommand{\smallcaption}[1]{\caption{\protect\small#1}}

\newcommand{\acronym}[3]{\newcommand{#1}{#3 (#2)\relax\renewcommand{#1}{#2}}}
\sizedef % settings for text height and width (see macrosquist.tex)

\newcommand\NS{N_{\rm Schw}} % Schwarzschild radial gaugefunction

\newcommand\approxless{\lesssim}
\newcommand\approxlarger{\gtrsim}

\acronym{\hw}{HW}{{Harrison-Wheeler}}
\acronym{\gbone}{GB1}{{\em generalized Buchdahl $n=1$ polytrope}}
\acronym{\gbfive}{GB5}{{\em generalized Buchdahl $n=5$ polytrope}}
\acronym{\sss}{SSS}{{\em static spherically symmetric}}

\begin{document}

\reportnumber{SUITP 98-20}
\eprintnumber{gr-qc/9809055}

\title{Exact relativistic stellar models with liquid surface. \\
       I. Generalizing Buchdahl's $n=1$ polytrope.}
\author{Kjell Rosquist \\
        {\small Department of Physics, Stockholm University}  \\[-10pt]
        {\small Box 6730, 113 85 Stockholm, Sweden} \\
        {\small E-mail: \textsf{kr@physto.se}}}
\date{}
\maketitle

\begin{abstract}
A family of exact relativistic stellar models is described.  The family 
generalizes Buchdahl's $n=1$ polytropic solution.  The matter content is a 
perfect fluid and, excluding Buchdahl's original model, it behaves as a liquid 
at low pressures in the sense that the energy density is non-zero in the zero 
pressure limit.  The equation of state has two free parameters, a scaling and 
a stiffness parameter.  Depending on the value of the stiffness parameter the 
fluid behaviour can be divided in four different types.  Physical quantities 
such as masses, radii and surface redshifts as well as density and pressure 
profiles are calculated and displayed graphically.  Leaving the details to a 
later publication, it is noted that one of the equation of state types can 
quite accurately approximate the equation of state of real cold matter in the 
outer regions of neutron stars.  Finally, it is observed that the given 
equation of state does not admit models with a conical singularity at the 
center.
\end{abstract}

%\centerline{\bigskip\noindent PACS numbers: \gr \quad \cosmology}
%\clearpage

%%%%%%%%%%%%%%%%%%%%%%%%%%%%%%%%%%%%%%%%%%%%%%%%%%%%%%%%%%%%%%%%%%%%%%%%%%%%%%
\section{Introduction}
%%%%%%%%%%%%%%%%%%%%%%%%%%%%%%%%%%%%%%%%%%%%%%%%%%%%%%%%%%%%%%%%%%%%%%%%%%%%%%

Among the new exact relativistic stellar models recently found in 
\cite{rosquist:star,simon:polyfive}, two families are of particular physical 
interest.  They both generalize gaseous models found by Buchdahl 
\cite{buchdahl:polyone,buchdahl:polyfive}.  His models have equations of state 
which reduce to a Newtonian polytropic gas (of index $n=1$ and $n=5$ 
respectively) at low pressures.  We will refer to the generalized solutions as 
the \gbone\ and \gbfive\ models respectively.  By contrast to Buchdal's 
original solutions the generalized models have equations of state which are 
liquid at low pressures in the sense that the density is non-zero at zero 
pressure, $\rho_{\rm s}>0$.\footnote{We use a subindex ``s'' throughout the 
paper to denote the surface values (where $p=0$) of various physical 
quantities.} Their equations of state each have two free parameters.  By 
properly adjusting those parameters, the models can describe the interior of 
neutron stars with surprising accuracy.  In this paper we describe the \gbone\ 
family \cite{rosquist:star}.  The matter in the model is a perfect fluid 
obeying a certain barotropic equation of state $p=p(\rho)$.  The history of 
\sss\ models with matter goes back to Schwarzschild's discovery already in 
1916 of his interior solution with incompressible matter 
\cite{schwarzschild:interior}.  Although a number of exact \sss\ models have 
been found since then (see {\em e.g.} Tolman's classic paper 
\cite{tolman:sssfluids}) there is an important distinction which should be 
made between solutions such as Schwarzschild's interior solution and solutions 
of Tolman's type.  Whereas the former kind is a family of solutions describing 
a stellar sequence with varying masses for the given equation of state ($\rho 
= constant$ in that case) the latter kind gives only a single solution with a 
fixed mass once the equation of state is specified.  The underlying reason is 
that Schwarzschild's interior model is the general solution of the Einstein 
field equations for the given equation of state.  By general solution we mean 
a solution with the full number of integration constants.  For any given 
equation of state there is a family of \sss\ models parametrized by precisely 
two physically relevant integration constants.  Usually one of these constants 
is determined by requiring that the stellar model is regular at the center.  
The single remaining nontrivial integration constant then determines the total 
mass.

The general solutions correspond to nontrivial symmetries of the field 
equations.  Such symmetries only exist for certain special equations of state 
\cite{rosquist:star}.  Tolman type solutions on the other hand are particular 
solutions which do not contain the full number of integration constants.  Such 
special solutions have been referred to as submanifold solutions since they 
correspond to submanifolds in the phase space which are left invariant by the 
field equations \cite{ujr:hh}.  Obviously the general solutions are more 
useful than the submanifold solutions.  However, until recently 
Schwarzschild's interior solution was the only general \sss\ solution which 
was considered to be of physical interest.

 At this point we should make clear what we mean by a physically interesting 
 solution.  For the purposes of this paper we consider a solution to be 
 physical if the energy density is non-negative, $\rho\geq0$ (gravity is 
 attractive), and that the matter is locally mechanically stable so that 
 $p\geq0$ and $dp/d\rho\geq0$.  Incompressible matter is considered to be 
 included in this category.  Schwarzschild's interior solution was the first 
 matter filled general physical \sss\ solution of the Einstein equations.  It 
 took almost half a century before the next general physical \sss\ model, the 
 $n=5$ polytrope solution, was discovered by Buchdahl in 1964 
 \cite{buchdahl:polyfive}.  As in the case of a Newtonian $n=5$ polytrope this 
 solution has an infinite radius although the mass is finite.  This model was 
 not considered to be of any significant physical interest.  A few years later 
 Buchdahl found the $n=1$ polytrope solution \cite{buchdahl:polyone}.  
 Although this model has a finite radius it has not been widely discussed in 
 the literature.  The next development came recently when the generalizations 
 of Buchdahl's solutions were discovered by Simon \cite{simon:polyfive} and 
 Rosquist \cite{rosquist:star}.  The \gbone\ and \gbfive\ solutions are in 
 fact families of solutions with equations of state depending on two 
 continuous parameters.  In the present paper we discuss the \gbone\ family in 
 some detail.  Because of the two parameters the structure of this family is 
 quite rich with qualitative behaviour varying in different parts of parameter 
 space.  In particular, by an appropriate choice of the equation of state 
 parameters, the \gbone\ family can be fitted quite accurately to the \hw\ 
 equation of state \cite{htww:gravcollapse} for cold matter expected to be 
 valid in the outer parts of a neutron star.  Later refinements of the \hw\
 equation of state are essentially corrections in the regime above nuclear 
 densities.  The fitting will be performed in a separate paper 
 \cite{rosquist:gb1b}.  It turns out that the best fit \gbone\ model is very 
 close to the original Buchdahl solution rendering that model more interesting 
 than previously thought.  As for the \gbfive\ family, its basic properties 
 are planned to be discussed in a forthcoming paper \cite{rosquist:gb5a} (see 
 also \cite{rosquist:trapped}).

One reason that the new static models are interesting is that they can serve 
as starting points to find an interior solution for the Kerr metric.  Assuming 
that such an interior general physical solution exists, and that it is valid 
for a slowly rotating Kerr exterior, there must be an \sss\ interior solution 
in the limit of zero rotation.  Furthermore, this static limit will also be a 
general physical solution because of the arbitrary mass parameter which is 
present in the Kerr metric.  Of course this heuristic argument excludes 
interiors with singular matter distributions such as the Negebauer-Meinel disk 
\cite{nm:disk}.  Referring to the extensive classification of general \sss\ 
models in \cite{rosquist:star}, the \gbone\ and \gbfive\ solutions are in fact 
the only good candidates for the static limit of a possible interior Kerr 
solution.  On the other hand it is quite conceivable that there are rotating 
generalizations of the \gbone\ and \gbfive\ models (as of any fluid \sss\ 
model) which have an exterior vacuum field which is different from Kerr.  One 
might try to spin up the \gbone\ model to yield a rotating perfect fluid 
solution which may or may not have an exterior gravitational field which 
coincides with the Kerr solution.  Attempts in this direction have been made 
by trying to extend the use of the Newman-Janis trick \cite{nj:trick} to 
impart rotation to fluid \sss\ models \cite{hj:newmanjanis,dt:newmanjanis} but 
without success so far.  However, the \gbone\ or \gbfive\ models so far have 
not been considered in this context.

Exact \sss\ solutions may also serve as benchmarks to test the numerical codes 
for integrating the equations of stellar structure.  Such testing was 
performed for example in \cite{ab:neutronstar} using the exact solution for 
the $n=1$ Newtonian polytrope.  The existence of exact relativistic solutions 
with physically reasonable equations of state make it possible to test such 
codes also in the relativistic domain.

Another important use for exact solutions is to investigate in more detail how 
the maximum mass depends on properties of the equation of state.  Only general 
physical solutions qualify as objects to study in this context since it is 
necessary to vary the mass while keeping the equation of state fixed.  
However, in the past there was no exact model available which was capable of 
exhibiting a mass maximum.  The reason is that the mass maximum requires a 
rather soft equation of state by contrast to the hardest conceivable matter, 
the incompressible perfect fluid, which makes up the matter source for the 
interior Schwarzschild solution.  The \gbone\ models which are the subject of 
this paper also contain matter which is too hard for a mass maximum to appear.  
However, as will be shown in \cite{rosquist:gb5a}, the \gbfive\ models have a 
softer equation of state and do exhibit a mass maximum.

In addition to the topics discussed above the \gbone\ models have also turned 
out to be interesting from the point of view of understanding the space of 
solutions of the \sss\ Einstein equations.  Taking the case of an 
incompressible fluid as an example, one of the two physically relevant 
integration constants is determined by the requirement that the metric should 
be regular at the center.  This is sometimes referred to as the condition of 
elementary flatness.  The center is defined as that value of the radial 
parameter for which the area of the 2-sphere of constant radius tends to zero.  
However, as described in section \ref{sec:physical} the \gbone\ models have a 
different behaviour in this respect.  It turns out that a \gbone\ model can 
only be continued to the center for one specific value of one of the 
integration constants.  The condition of elementary flatness is then 
automatically satisfied.  The global behaviour of singular \gbone\ models 
therefore seem to be different from that of singular incompressible models.  A 
discussion of possible singularity types for \sss\ models was given in 
\cite{wheeler_james:ssssing} but we have not made an attempt to see how the 
\gbone\ models would fit in that scheme.

%%%%%%%%%%%%%%%%%%%%%%%%%%%%%%%%%%%%%%%%%%%%%%%%%%%%%%%%%%%%%%%%%%%%%%%%%%%%%%
\section{The model}
%%%%%%%%%%%%%%%%%%%%%%%%%%%%%%%%%%%%%%%%%%%%%%%%%%%%%%%%%%%%%%%%%%%%%%%%%%%%%%
The usual starting point when studying \sss\ models is to write down the 
metric in Schwarzschild coordinates
\begin{equation}
   ds^2 = -\e^{2\nu}dt^2 + \e^{2\lambda}dr^2 + r^2 d\Omega^2 \ ,
\end{equation}
where
\begin{equation}
   d\Omega^2 = d\theta^2 + \sin^2\!\theta \,d\phi^2 \ ,
\end{equation}
is the metric of the 2-sphere and $\nu$ and $\lambda$ are functions of 
Schwarzschild's radial variable, $r$.  However, this form of the metric is not 
general enough in the sense that it is not possible to express all exact 
solutions in closed form using Schwarzschild coordinates.  It is therefore 
useful to write the \sss\ metric in terms of a general radial variable, $R$,
\begin{equation}\label{eq:genmetric}
   ds^2 = -\e^{2\nu}dt^2 + N^2dR^2 + W^2 d\Omega^2 \ ,
\end{equation}
where $N$ and $W$ are both functions of $R$.  Schwarzschild's radial variable 
then becomes a derived quantity given by $r=W(R)$.  The form of the metric 
\eqref{eq:genmetric} reflects the free gauge choice of the radial variable  
expressed by the freedom to choose the radial gauge function $N$.
To write down the \gbone\ model we use the gauge $N = \e^{-\nu}$. The \sss\ 
metric then becomes
\begin{equation}
     ds^2 = - Z dt^2 + Z^{-1} dR^2 + W^2 d\Omega^2 \ ,
\end{equation}
where $W$ and $Z := \e^{2\nu}$ are functions of the radial parameter $R$.  
Since $N>0$, the radial variable $R$ is increasing outwards from the center of 
the star.  The Schwarzschild radial variable is given by $r=W(R)$.  The center 
of the star\footnote{We use the subindex ``c'' to denote the center of the 
star.} is therefore located at $R=R_{\rm c}$ defined by the condition 
$W(R_{\rm c}) =0$.  The \gbone\ model is defined by \cite{rosquist:star} 
\footnote{For convenience we have performed the rescalings $t \rightarrow 4t$, 
$ds^2 \rightarrow 16ds^2$, $T \rightarrow 4T$ and $X \rightarrow 4X$ relative 
to the conventions of reference \cite{rosquist:star}.}
\begin{equation}\label{eq:solution}
 \begin{split}
     &Z = V/W \ ,\qquad W = T + X \ ,\qquad V = T - X \ ,\\
     &T = \begin{cases}
          (\omega_-)^{-1} (\cosh\zeta) \sin[\omega_-(R-R_-)] \ ,
                                    & \text{for $\delta > 1$ \ ,} \\
          (\cosh\zeta) (R-R_-)      & \text{for $\delta = 1$ \ ,}
          \end{cases} \\
     &X = (\omega_+)^{-1} (\sinh\zeta) \sin[\omega_+(R-R_+)] \ ,
 \end{split}
\end{equation}
where $\omega_\pm = \sqrt{2\kappa a(\delta\pm1)}$ and $\zeta>0$.  The 
parameters $\zeta$ and $R_\pm$ are integration constants while the parameters 
$a$ and $\delta$ characterize the equation of state.  The latter can be 
parametrized as
\begin{equation}\label{eq:eqstate}
    p = a(Z^2 - 2\delta Z + 1) \ ,\qquad \rho = a(-5Z^2 + 6\delta Z - 1) \ ,
\end{equation}
where $a>0$ and $\delta\geq1$.  As discussed in \cite{rosquist:star}, models 
with $\delta <1$ also exist but can only be used in the pressure region $p 
\geq a(1-\delta^2)$.  The range of $\delta$ will be further restricted below.  
From the physical point of view, $a$ is a scaling parameter.  Models with 
different values of $a$ can be transformed into each other by simply scaling 
the metric.  The parameter $a$ represents an overall scaling of length scales 
in the model.  A model with given $a$ has a natural length scale given by 
$\ell = (\kappa a)^{-1/2}$.  However, scale invariant physical quantities such 
as surface redshift are unaffected by changes in $a$.  As will be explained 
below the parameter $\delta$ can be interpreted as a measure of the stiffness 
of the matter.  We will refer to it as the stiffness parameter.  The 
integration constants are not all physical.  Only $\zeta$ and the difference 
$D := R_+ - R_-$ carry physical information.  This is because $D$ is invariant 
under translations of $R$.  Because of its nature as a metric coefficient $Z$ 
must be strictly positive.  Physical requirements on the equation of state 
further restrict the range of $Z$ as discussed in the next section.

Expressing the functions $T$ and $X$ in terms of $r$ and $Z$ we have
\begin{equation}\label{eq:TXineq}
   T = \case12 r(1+Z) \ ,\qquad X = \case12 r(1-Z) \ .
\end{equation}
Since the conditions $r \geq 0$ and $Z>0$ must hold in the stellar interior
it  follows that $T \geq 0$.  Also, as shown below in \eqref{eq:Zcond3}, we
have  that $Z \leq 1$ which shows that $X \geq 0$.

%%%%%%%%%%%%%%%%%%%%%%%%%%%%%%%%%%%%%%%%%%%%%%%%%%%%%%%%%%%%%%%%%%%%%%%%%%%%%%
\subsection{The equation of state}
%%%%%%%%%%%%%%%%%%%%%%%%%%%%%%%%%%%%%%%%%%%%%%%%%%%%%%%%%%%%%%%%%%%%%%%%%%%%%%
In this paper we assume that the energy density is non-negative, $\rho\geq0$, 
and that the matter is mechanically stable so that $p\geq0$ and 
$dp/d\rho\geq0$.  It follows immediately from (\ref{eq:eqstate}) that the 
condition $p \geq 0$ is satisfied precisely if
\begin{equation}\label{eq:Zcond2}
     Z \leq Z_{\rm s} = \delta - \sqrt{\delta^2-1} \ ,
\end{equation}
where $Z_{\rm s}$ denotes the value of $Z$ at the star surface defined by the 
condition $p=0$. Also, the relation 
\begin{equation}\label{eq:dpdrho}      
     \frac{dp}{d\rho} = \frac{\delta-Z}{5Z-3\delta} \ .
\end{equation}
together with $Z>0$ implies that the condition $dp/d\rho\geq0$ is equivalent 
to the relation
\begin{equation}\label{eq:Zcond1}
     \frac{3\delta}{5} < Z \leq \delta \ .
\end{equation}
The speed of sound is given by $u := \sqrt{dp/d\rho}$.  It reaches the speed 
of light, $u=c=1$, at $Z_*=2\delta/3$.  It is sometimes of interest to allow 
superluminal sound speeds for example to obtain a simple model such as 
Schwarzschild's interior solution.  The limit in which the speed of sound 
becomes superluminal will be referred to as the causal limit and denoted by an 
asterisk subscript.  The pressure at the causal limit is given by $p_* = 
a(1-8\delta^2/9)$ and the corresponding energy density is $\rho_* = 
a(16\delta^2/9-1)$.  At zero pressure the energy density is $\rho_{\rm s} = 
4a(1 - \delta^2 + \delta \sqrt{\delta^2-1})$.  The speed of sound at the 
surface is given by
\begin{equation}
     u_{\rm s}^2 = \frac{\sqrt{\delta^2-1}}{2\delta-5\sqrt{\delta^2-1}} \ .
\end{equation}
It follows from this expression that the zero pressure sound speed reaches the 
speed of light at $\delta = \delta_* := 3\sqrt2 /4 \approx 1.061$ and becomes 
infinite in the limit\footnote{We denote the limit of infinite sound speed by 
a subscripted infinity symbol.} $\delta \rightarrow \delta_\infty := 
5/\sqrt{21} \approx 1.091$.  Thus models with $\delta>\delta_*$ have a 
superluminal speed of sound for all physcially allowed values of the pressure 
while models with $\delta\leq \delta_*$ are causal for $p<p_*$ and acausal or 
$p>p_*$.

The upper limit in (\ref{eq:Zcond2}) must be strictly larger than the lower 
limit in (\ref{eq:Zcond1}).  This gives the additional restriction $\delta < 
\delta_\infty$ and then
\begin{equation}\label{eq:Zcond3}
     \frac{3\delta}{5} < Z \leq \delta - \sqrt{\delta^2-1} \leq 1 \ .
\end{equation}
It can be checked that the condition $\rho \geq 0$ is always satisfied if
$\delta \geq 1$ and $Z$ is in the range (\ref{eq:Zcond3}).

An important physical characteristic of the equation of state is the 
relativistic adiabatic index defined by
\begin{equation}
     \gamma = \left(1+\frac{\rho}{p}\right) \frac{dp}{d\rho} \ .
\end{equation}
It can be interpreted physically as a compressibility index.  Lower 
$\gamma$-values correspond to softer matter.  For the \gbone\ model the 
adiabatic index is given by
\begin{equation}
     \gamma = \frac{4Z(\delta-Z)^2}{(5Z-3\delta)(Z^2-2\delta Z+1)} \ .
\end{equation}
For models with $\delta>1$ the adiabatic index tends to infinity at the 
surface while $\gamma_{\rm s}=2$ for Buchdahl's original model ($\delta=1$).  
The adiabatic index also tends to infinity in the limit $Z \rightarrow 
Z_\infty := 3\delta/5$ corresponding to $v_{\rm sound} \rightarrow \infty$.  
At the causal limit the adiabatic index is given by $\gamma_* = 
8\delta^2/(9-8\delta^2)$.

In practical calculations it is convenient to use $\xi := \omega_-/\omega_+ = 
\sqrt{(\delta-1) /(\delta+1)}$ as a stiffness parameter in place of $\delta$.  
The value $\xi=0$ (Buchdahl's original model) corresponds to the softest 
equation of state within the \gbone\ family.  The range of $\xi$ is
\begin{equation}
     0 \leq \xi < \xi_\infty := \frac12(5-\sqrt{21}) \approx 0.2087 \ .
\end{equation}
The causal limit is at $\xi = \xi_* := 3-2\sqrt2 \approx 0.1716$.  The 
parameter $\xi$ can be interpreted physically in terms of the speed of sound.  
Using \eqref{eq:eqstate} and \eqref{eq:dpdrho} and defining $k:=p/\rho$ we 
find that
\begin{subequations}\label{eq:xi}
 \begin{align}
   \xi &= \frac{\sqrt{d}}{\sqrt{d_1}+\sqrt{d_2}} \ ,\\
  \intertext{where}
   d &:= d_1 - d_2 = 4u^2(u^2-2k-5u^2k) \ ,\\
   d_1 &:= (k+1)(1+5u^2)^2 \ ,\\
   d_2 &:= k+1 + 2(9k+5)u^2 + 3(15k+7)u^4 \ .
 \end{align}
\end{subequations}
In the limit $k \rightarrow 0$ this reduces to
\begin{equation}
   \xi = \frac{2u_{\rm s}^2}{1+5u_{\rm s}^2 
                                     + \sqrt{1+10u_{\rm s}^2+21u_{\rm s}^4}}
       = u_{\rm s}^2 + \mathcal{O}(u_{\rm s}^4) \ .
\end{equation}
The inverse of this relation takes the simple form $u_{\rm s}^2 = \xi/\mu$ 
where we have defined the parameter $\mu := 1-5\xi+\xi^2$ which takes values 
in the interval $0< \mu \leq1$.  Thus for $u_{\rm s}\ll1$ we see that $\xi$ is 
essentially the sound speed squared.  Assuming that the material in a cold 
star at low pressure consists of pure iron we can get a rough estimate of the 
value of $\xi$ for a realistic stellar model.  The speed of sound in iron is 
$6\cdot10^5 cm/s$ \cite{crc:handbook73} corresponding to $u=2\cdot 10^{-5}$ in 
geometrized units.  Inserting this value in \eqref{eq:xi} gives $\xi \sim 
4\cdot10^{-10}$.  This value is not sensitive to the value of $p$ for moderate 
pressures up to and beyond $p=1$ atm.  In \cite{rosquist:gb1b} we will 
determine $\xi$ more precisely by fitting the \gbone\ equation of state to the 
\hw\ equation of state.  This will also involve a determination of the scaling 
parameter $a$.

%^^^^^^^^^^^^^^^^^^^^^^^^^^^^^^^^^^^^^^^^^^^^^^^^^^^^^^^^^^^^^^^^^^^^^^^^^^^^%
\begin{figure}[!tbp]
\centering
\includegraphics{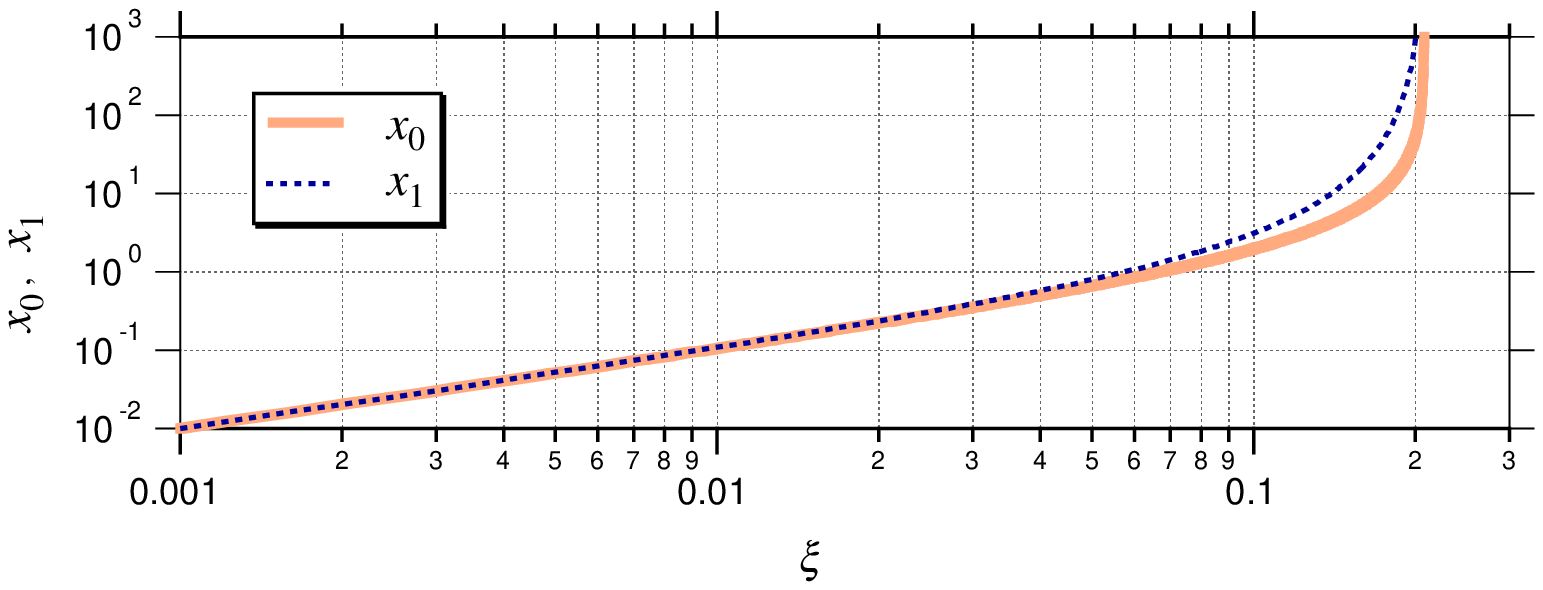}
\smallcaption{The two equation of state parameters $x_0$ and $x_1$ are plotted 
         against $\xi$ using logarithmic scales. Both $x_0$ and $x_1$ tend to
         infinity in the limit $\xi\rightarrow \xi_\infty \approx 0.2087$.
         They are both approximately unity at $\xi = 0.06$.}
\label{fig:x0x1}
\end{figure}
\begin{figure}[!tbp]
\centering
\psfrag{prho_left}[][]{\Large $\hat p, \hat\rho$}
\psfrag{prho_bottom}[][]{\Large $x$}
\includegraphics{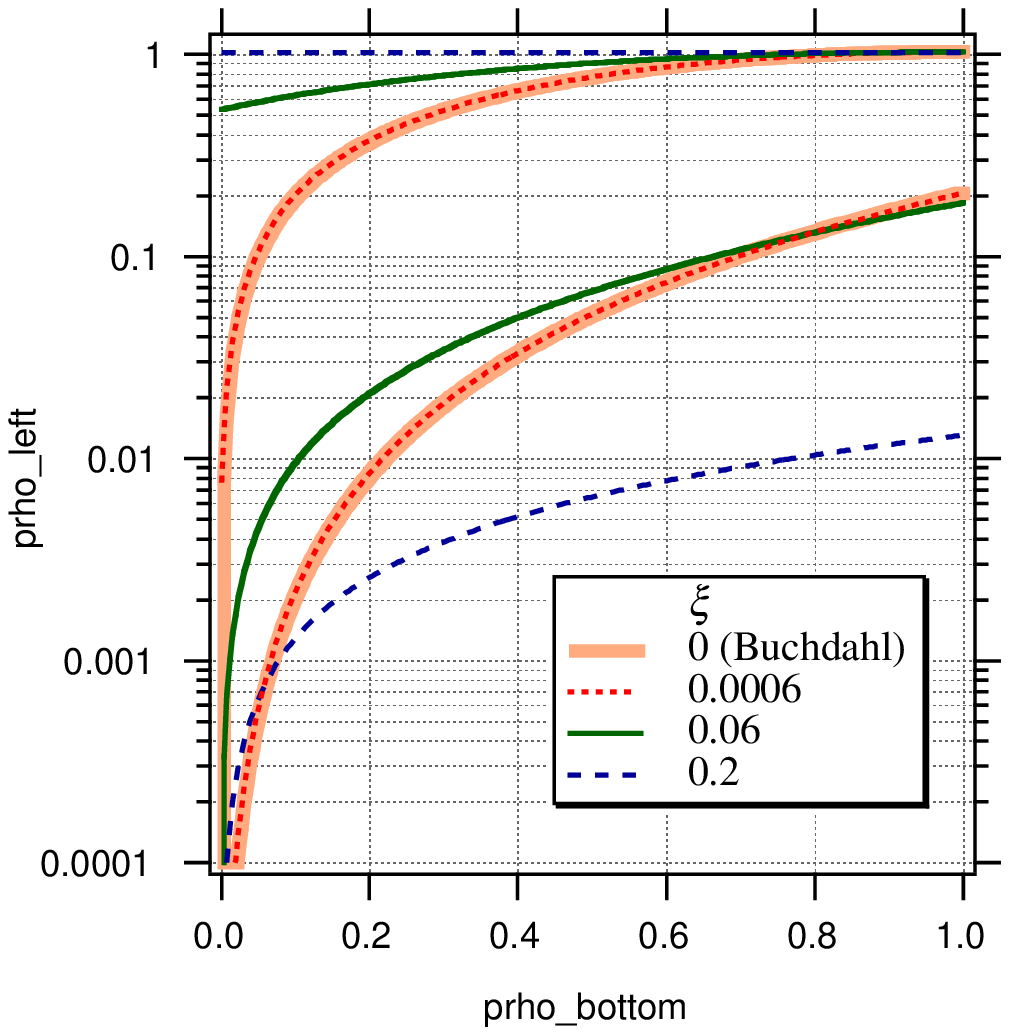}
     \smallcaption{ The pressure and the energy density are plotted for four 
     values of $\xi$ representing the different types of equation of state as 
     discussed in the text.  For each of these values the upper curve 
     corresponds to the normalized energy density while the lower curve gives 
     the normalized pressure.  \comment{The density curve for $\xi=0.0006$ 
     has the nonzero surface value $\hat\rho_{\rm s} \approx 0.0062$.} The 
     causal limit $\rho=\rho_*$ in this diagram is the line 
     $\hat\rho=1$.  }
\label{fig:prho}
\end{figure}
\begin{figure}[!tbp]
\centering
\psfrag{gamma_left}[][]{\Large $\gamma$}
\psfrag{gamma_bottom}[][]{\Large $\xi$}
\includegraphics{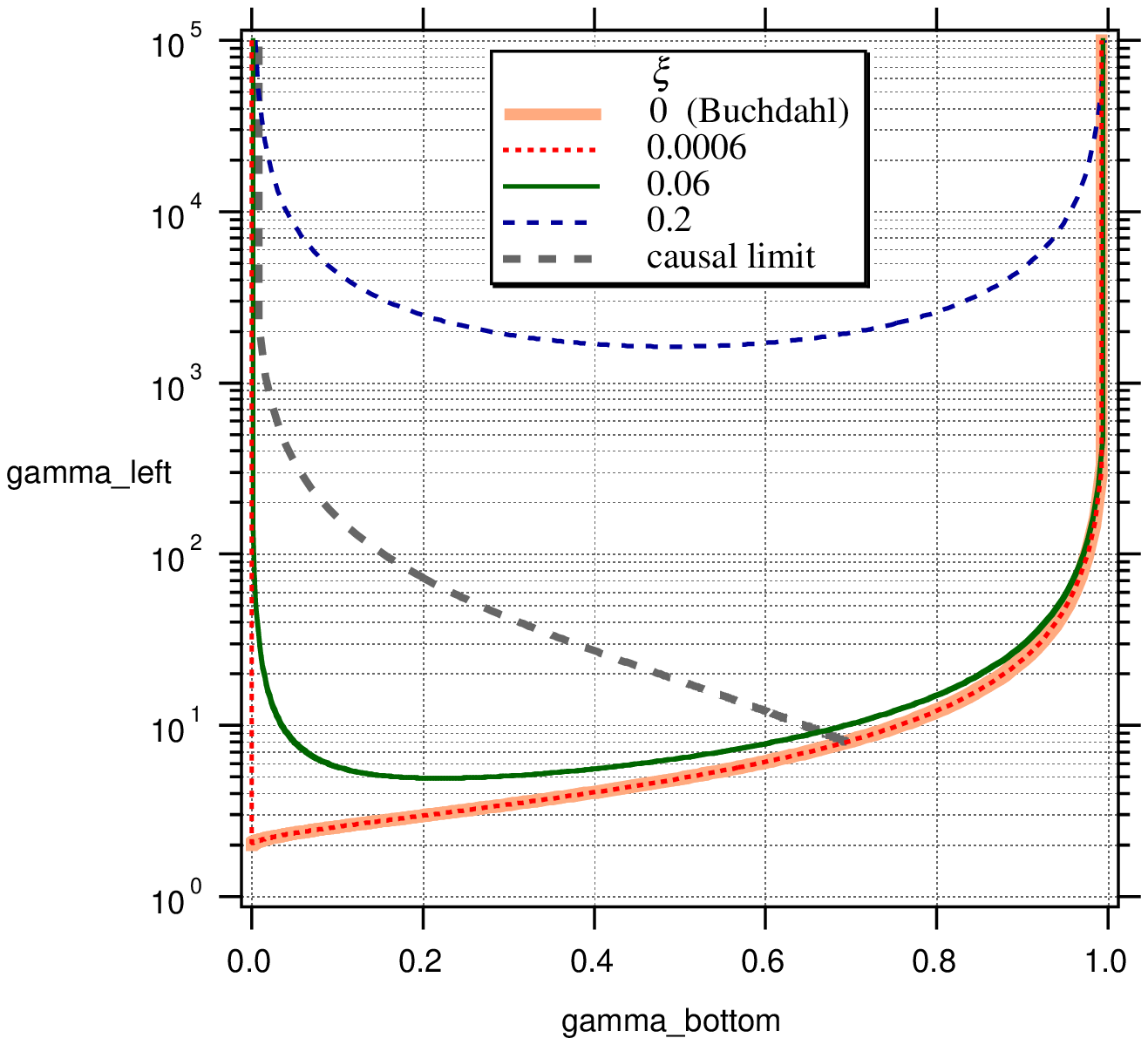}
\smallcaption{ The adiabatic index for the $n=1$ Buchdahl equation of state 
and some of its generalizations.  The pressure is scaled by the pressure 
$p_\infty$ at which the adiabatic index becomes infinite.  For $\xi>0$ the 
adiabatic index diverges in the zero pressure limit.  The indicated causal 
limit is the boundary between the regions with subluminal and superluminal 
sound velocity.  It is seen that the causal region defined by subluminal sound 
speed is located in the lower left hand corner of the diagram with a narrow 
section stretching towards infinity along the $\gamma$-axis.  Models with $\xi 
< \xi_* = 3-2\sqrt2 \approx 0.1716$ start in the causal region at low 
pressures and enter the acausal region at the pressure $p_*$ while the models 
with $\xi > \xi_*$ are acausal for all pressures.  }
\label{fig:gamma}
\end{figure}

%----------------------------------------------------------------------------%
It is also convenient to replace $Z$ by a new normalized parameter, $x$, which 
takes values in the interval [0,1] with $x_{\rm s}=0$ and $x_\infty=1$.  This 
is accomplished by the transformation $Z=Kx+Z_{\rm s}$ where $K:= Z_\infty 
-Z_{\rm s}= -(2/5)\mu/(1-\xi^2)$ and $Z_{\rm s}=(1-\xi)/(1+\xi)$ which implies
\begin{subequations}\label{eq:eqstate_x}
 \begin{align}\label{eq:eqstate_xa}
      p &=  b x (x + x_0) \ ,\\
  \label{eq:eqstate_xb}
   \rho &= 5b (-x^2 + 2x + x_1) \ ,
 \end{align}
\end{subequations}
where $b=aK^2$ and
\begin{equation}
   x_0 = 10\xi\mu^{-1} \ ,\quad
   x_1 = 10\xi(1-\xi)^2\mu^{-2} \ .
\end{equation}
These constants satisfy the inequality $0\leq x_0\leq x_1$.  They are plotted 
as functions of $\xi$ in figure \ref{fig:x0x1}.  We see that $x_0$ and $x_1$ 
are approximately equal up to $\xi \sim 0.06$ corresponding to $x_0,x_1 
\approxless 1$.  In the expression for the pressure \eqref{eq:eqstate_xa} the 
term $x_0$ comes into play if $x_0 \approxlarger x$.  Similarly in the 
expression for the energy density \eqref{eq:eqstate_xb} the term $x_1$ is 
important if $x_1 \approxlarger x(2-x)$.  Since the functions $x$ and $x(2-x)$ 
both take values in the interval $[0,1]$, this leads to a rough subdivision of 
the \gbone\ equations of state into the four types
\begin{equation}
   \begin{array}{rll}
 \mbox{(I)}   & x_0=x_1=0       & (\xi=0) \ ,\\
 \mbox{(II)}  & 0< x_0,x_1 \ll1 & (0<\xi \approxless 0.001) \ ,\\
 \mbox{(III)} & x_0,x_1\sim1   & (0.001\approxless\xi\approxless 0.2)\ ,\\
 \mbox{(IV)}  & x_0,x_1\gg1     & (0.2\approxless\xi<\xi_\infty\approx0.2087).
    \end{array}
\end{equation}
As discussed above, comparison with the actual equation of state of cold 
matter at low pressure demands $\xi \sim 4\cdot10^{-10}$ corresponding to a 
type II equation of state.  The four different types of equation of state are 
plotted in figures \ref{fig:prho} and \ref{fig:gamma}.  The type I and II 
equations of state are practically indistinguishable in figure \ref{fig:prho} 
but differ drastically in the behaviour of $\gamma$ in the zero pressure limit 
as illustrated in figure \ref{fig:gamma}.

%^^^^^^^^^^^^^^^^^^^^^^^^^^^^^^^^^^^^^^^^^^^^^^^^^^^^^^^^^^^^^^^^^^^^^^^^^^^^%
%----------------------------------------------------------------------------%

%^^^^^^^^^^^^^^^^^^^^^^^^^^^^^^^^^^^^^^^^^^^^^^^^^^^^^^^^^^^^^^^^^^^^^^^^^^^^%
%----------------------------------------------------------------------------%

In figure \ref{fig:prho} the pressure and energy density are plotted as 
functions of $x$ and normalized by the energy density at the causal limit 
which occurs at
\begin{equation}
     x = x_* := \frac{5(1-6\xi+\xi^2)}{6(1-5\xi+\xi^2)} \ .
\end{equation}
The energy density and pressure in that limit are given by
\begin{equation}
   \rho_* = \case{25}{36}b\mu^{-2}(7+\xi^2)(1+7\xi^2) \ ,\qquad
      p_* = \case{25}{36}b\mu^{-2}(1-34\xi^2+\xi^4) \ .
\end{equation}
Since any given value of $\rho$ corresponds dimensionally to a pressure $\rho 
c^2$ it is reasonable to normalize both the energy density and the pressure by 
$\rho_*$.  The dimensionless normalized pressure and energy density then 
become
\begin{equation}
 \begin{split}
      \hat p := p/\rho_* &= \hat b \,x(x + x_0) \ ,\\
 \hat\rho := \rho/\rho_* &= 5\hat b (-x^2+2x+x_1) \ ,
 \end{split}
\end{equation}
where
\begin{equation}
   \hat b := \frac{36(1-5\xi+\xi^2)^2}{25(7+\xi^2)(1+7\xi^2)} \ .
\end{equation}
To see the full picture when plotting $\gamma$ we use the pressure 
$p_\infty$ at which $\gamma$ tends to infinity as a reference point.  Setting 
$x=1$ in \eqref{eq:eqstate_xa} gives
\begin{equation}
   p_\infty = b\cdot\frac{1+5\xi+\xi^2}{1-5\xi+\xi^2} \ .
\end{equation}
Then we have
\begin{equation}
   p/p_\infty = \tilde b\,x(x + x_0) \ ,
\end{equation}
where
\begin{equation}
   \tilde b = \frac{1-5\xi+\xi^2}{1+5\xi+\xi^2} \ .
\end{equation}
Writing the adiabatic index in terms of the normalized parameter gives
\begin{equation}
   \gamma = \frac{2(5\xi + \mu x)^2 [\mu(5-2x)+15\xi]}
                 {5\mu^2 x(1-x)(10\xi+\mu x)} \ .
\end{equation}
For Buchdahl's polytrope this reduces to
\begin{equation}
   \gamma = \frac{2(5-2x)}{5(1-x)} \ .
\end{equation}
At zero pressure we have $\gamma=2$ for this model.  For the models with 
$\xi>0$ the adiabatic index instead tends to infinity in that limit with the 
asymptotic form $\gamma \approx \rho_{\rm s} u_{\rm s}^2 p^{-1}$.

%%%%%%%%%%%%%%%%%%%%%%%%%%%%%%%%%%%%%%%%%%%%%%%%%%%%%%%%%%%%%%%%%%%%%%%%%%%%%%
\section{Physical characteristics of the stellar model}
\label{sec:physical}
%%%%%%%%%%%%%%%%%%%%%%%%%%%%%%%%%%%%%%%%%%%%%%%%%%%%%%%%%%%%%%%%%%%%%%%%%%%%%%
In this section we compute some of the general physical characteristics of the 
\gbone\ family.  As already stated the actual fitting to the \hw\ equation of 
state will be done in the sequel paper \cite{rosquist:gb1b}.  First we briefly 
discuss criteria which determine the degree to which the star may considered 
as relativistic.  One such criterion is the surface redshift $z_{\rm s}$ which 
in principle can be measured directly.  In practice, however, this seems to be 
unrealistic for neutron stars.  However, redshifts can be determined 
indirectly if the mass and radius of the star is known.  The surface redshift 
is given by the formula $z_{\rm s} = (1-2\beta)^{-1/2} -1$ where $\beta := 
M/r_{\rm s}$ (see {\em e.g.} \cite{schutz:gr}).  A given value of the redshift 
corresponds to an equivalent velocity through the Doppler relation $1+z= 
\sqrt{(1+v/c)/(1-v/c)}$.  For a Newtonian star, $z_{\rm s}\ll1$.  Defining a 
star to be relativistic if the surface redshift corresponds to a velocity 
greater than one percent of the speed of light this would then be the case if 
$z \approxlarger 0.01$.  In principle the redshift can attain values from the 
Newtonian $z_{\rm s} =0$ up to Buchdahl's limit \cite{buchdahl:limit} $z_{\rm 
s} =2$ (corresponding to $r_{\rm s} > \case94 M$) which is the maximum 
redshift if the density is assumed to be non-increasing in the outward 
direction, $d\rho/dr \leq 0$.  In practice, however, the surface redshift of a 
neutron star is expected to be in the range $0.1 \approxless z_{\rm s} 
\approxless 0.9$ \cite{lindblom:redshift}.  Another criterion can be 
formulated by considering the deviation from Euclidean geometry.  This 
deviation can be defined as the quantity $\alpha := 1-r/R_{\rm prop}$ where 
$R_{\rm prop}$ is the proper radial distance defined for a general \sss\ 
metric by $R_{\rm prop} = \int_{R_{\rm c}} N\d R$.  The star can then be said 
to be relativistic if, say, $\alpha_{\rm s} \approxlarger 0.01$.

Regular stellar models must satisfy the condition for elementary flatness at 
the center.  This can be conveniently expressed in terms of the Schwarzschild 
radial gauge function which for the \gbone\ models is given by 
\cite{rosquist:star}
\begin{equation}\label{eq:NSchw}
     \NS = Z^{-1/2} (W_{,R})^{-1} \ . 
\end{equation}
The elementary flatness condition then takes the form
\begin{equation}
   \lim_{r\rightarrow0} \NS = 1 \ .
\end{equation}
Using \eqref{eq:NSchw} it can be written as $Z_{\rm c}^{1/2} (W_{,R})_{\rm c} 
=1$.  From \eqref{eq:solution} we find $(X_{,R})_{\rm c} = \cosh\zeta$, 
$(X_{,R})_{\rm c} = \cosh\zeta$.  This implies $(W_{,R})_{\rm c} = \e^\zeta$, 
$(V_{,R})_{\rm c} = \e^{-\zeta}$ and $Z_{\rm c} = (V/W)_{\rm c} = 
(V_{,R})_{\rm c}/(W_{,R})_{\rm c} = \e^{-2\zeta}$.

In the remainder of this paper we will assume that the \gbone\ equation of 
state is valid throughout the star.  Since $Z$ is bounded it then follows from 
\eqref{eq:TXineq} that $T$ and $X$ both vanish at the center of the star.  
Using the translational freedom in $R$ we may set $R_{\rm c}=0$.  The 
condition $T(0)=0$ then implies $R_-=0$ for $\xi=0$ and that $\omega_- R_-$ is 
a multiple of $\pi$ for $\xi>0$.  Moreover, to ensure that $T$ is positive as 
$R$ increases away from the center $\omega_- R_-$ must actually be an even 
multiple of $\pi$.  Similarly, the condition $X(0)=0$ implies that $\omega_+ 
R_+$ must also be an even multiple of $\pi$.  As already stated in 
\cite{rosquist:star} we may then set $R_\pm=0$ without loss of generality.

Let us pause for a moment to digest what is actually implied here.  We have 
the curious situation that the mere assumption that the model can be continued 
to the center of the star implies the vanishing of one of the essential 
integration constants ($D = R_+ - R_- =0$).  Technically this has 
happened because the function $W(R)$ which gives Schwarzschild's radial 
variable is built up by two non-negative terms, $r = W(R) = T(R)+X(R)$, which 
must both vanish separately at $R=0$.  Normally one would expect to use that 
integration constant to ensure the elementary flatness of the model at the 
center.  So what about elementary flatness?

  It follows that the model 
in fact satisfies the elementary flatness condition \cite{rosquist:star}.  It 
seems that the \gbone\ equation of state cannot support a conical singularity 
at the center.  This result sparks off a number of questions.  For example, 
what equations of state have this property of not allowing conical 
singularities?  Also, since a conical singularity is not allowed, in what way 
does the model break down if we let say $R_+$ be nonzero? The answer to the 
latter question is that the equation of state itself becomes unphysical in the 
sense that $u \rightarrow \infty$ at some value of the 
Schwarzschild radial variable, $r=r_\infty >0$.  At still smaller radii,
$r<r_\infty$,  the matter becomes mechanically unstable, $dp/d\rho <0$.  Thus
by allowing a  nonzero $R_+$ what happens is that the matter becomes
unphysical before any  breakdown in the geometry as we move towards lower
$r$-values.

The radius of the star is defined as $r_{\rm s} = W(R_{\rm s})$.  The star 
mass can be computed by the formula \cite{rosquist:star}
\begin{equation}
     M = \case12 \left[ W(1-\NS^{-2}) \right]_{\rm s} \ .
\end{equation}
This leads to the following expressions for the physical parameters $M$, 
$\beta$ and $z_{\rm s}$
\begin{equation}\label{eq:physparam}    
             M = \case12 r_{\rm s} \left[
                     1 - Z (W_{,R})^2 \right]_{\rm s} \ ,\qquad
         \beta = \case12 \left[
                     1 - Z (W_{,R})^2 \right]_{\rm s} \ ,\qquad
   1+z_{\rm s} = Z_{\rm s}^{-1/2} (W_{,R})_{\rm s}^{-1} \ .
\end{equation}

\begin{figure}[!tbp]
\centering
\includegraphics{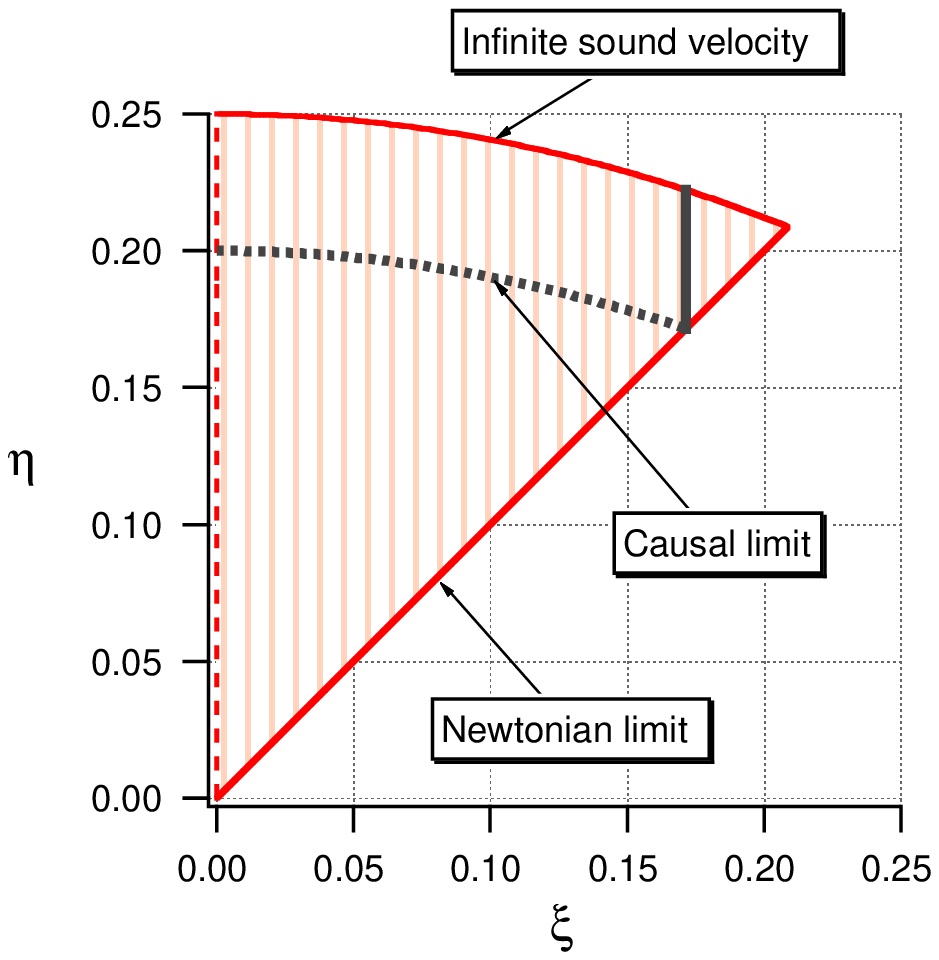}
\smallcaption{ The allowed $\xi-\eta$ region.  The vertical dashed line 
represents the original Buchdahl model ($\xi=0$).  The $\eta=\xi$ line 
corresponds to the low pressure Newtonian limit while the solid curve 
represents the most extreme relativistic models in the limit $u_{\rm c} 
\rightarrow \infty$.  From \eqref{eq:etalimits}, the solid curve is given by 
the equation $\eta=\eta_{\infty} = (1-4\xi^2)/(4-\xi^2)$.  The dotted line is 
the causal limit corresponding to models for which the speed of sound at the 
center equals the speed of light, $u_{\rm c} = 1$.  Its equation is $\eta = 
\eta_* := (1-5\xi^2)/(5-\xi^2)$.  Models lying below the causal limit are 
entirely causal while models above this limit have an acausal region in the 
stellar interior.  Models with $\xi> \xi_* = 3-2\sqrt2 \approx 0.1716$ (the 
region to the right of the vertical solid line) are acausal throughout the 
stellar interior.  For each given value of $\xi$ in the allowed range $0 \leq 
\xi < \xi_\infty \approx 0.2087$ there is a 1-parameter sequence of models 
represented by the vertical straight line which starts at the point $(\xi, 
\xi)$ and ends on the solid curve.  }
\label{fig:xi_eta_region}
\end{figure}

To evaluate these expressions at the surface we need to know $R_{\rm s}$.  
Using \eqref{eq:solution} and \eqref{eq:Zcond2} we find that $R_{\rm s} = 
q/\omega_+$ where $q$ is defined to be the smallest positive root of the 
equation $\sin(\xi x) = \eta \sin x$ where we have defined $\eta :=
\tanh\zeta$. Thus $q$ satisfies
\begin{equation}\label{eq:q}
     \sin(\xi q) = \eta \sin q \ .
\end{equation}
For Buchdahl's original model 
($\xi = 0$) we have $q = \pi$ and $R_{\rm s} = \pi/(2\sqrt{\kappa a})$.  For 
$\xi\neq0$ equation (\ref{eq:q}) cannot be solved explicitly.  Now what about 
the size of $\eta$ compared to $\xi$?  In fact $e^{-2 \zeta} = Z_{\rm c}$ is 
constrained by the requirement $Z_\infty < Z_{\rm c} < Z_{\rm s}$.  This leads 
to the inequality ({\it cf.} \cite{rosquist:star})
\begin{equation}
     3/5 \leq \frac{3(1+\xi^2)}{5(1-\xi^2)} = 3\delta/5 < e^{-2\zeta}
    < \delta-\sqrt{\delta^2-1} = \frac{1-\xi}{1+\xi} \leq 1 \ .
\end{equation}
Expressing this condition in terms of $\eta$ leads to
\begin{equation}\label{eq:etalimits}
    0 \leq \xi < \eta < \eta_\infty := \frac{1-4\xi^2}{4-\xi^2} \leq 1/4 \ .
\end{equation}
The allowed values of $\xi$ and $\eta$ are shown graphically in figure 
\ref{fig:xi_eta_region}.
%^^^^^^^^^^^^^^^^^^^^^^^^^^^^^^^^^^^^^^^^^^^^^^^^^^^^^^^^^^^^^^^^^^^^^^^^^^^^%
%----------------------------------------------------------------------------%

\begin{figure}[!tbp]
\centering
\includegraphics{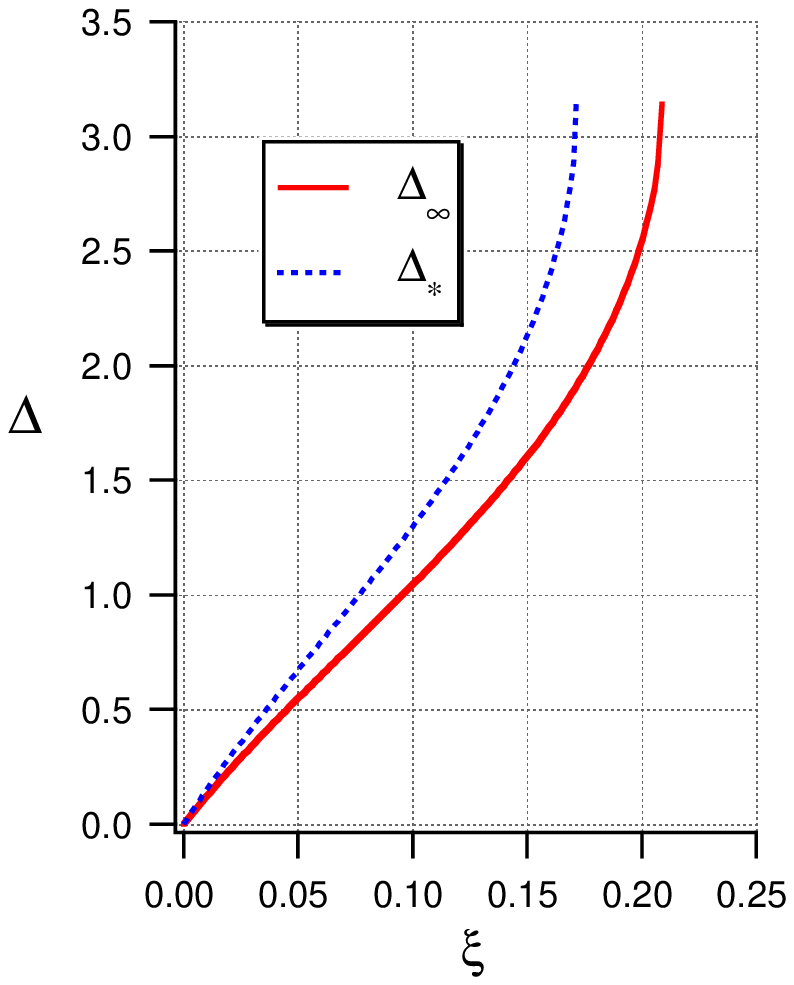}
\smallcaption{
The function $\Delta_\infty(\xi)$ is plotted in the interval $0\leq
\xi < \xi_\infty = (5-\sqrt{21})/2 \approx 0.2087$.  The function 
$\Delta_*(\xi)$ is plotted in the interval $0\leq \xi \leq \xi_* = 3-2\sqrt2 
\approx 0.1716$.  The models in the range $\xi_* < \xi < \xi_\infty$ have a
superluminal sound speed at the center for all values of $\eta$. }
\label{fig:delta}
\end{figure}

To determine the surface value of the radial coordinate we must solve equation 
(\ref{eq:q}) numerically for each set of values $(\xi, \eta)$ in the parameter 
space defined by \eqref{eq:etalimits}.  However, this can be avoided by 
observing that $\eta$ is an increasing function of $q$ for all $\xi$.  
Conversely, $q$ is an increasing function of $\eta$.  Therefore, we can use 
$q$ as a parameter in place of $\eta$ to represent the sequence of stellar 
models.  It follows that $q$ ranges from the minimal value $q = 0$ 
corresponding to $\eta = \xi$ (see (\ref{eq:etalimits})) to a certain 
maximal value $q = q_\infty$ corresponding to $\eta= \eta_\infty = 
(1-4\xi^2)/ (4-\xi^2)$.  The value $q_\infty$ can only be determined 
implicitly by the equation $\sin(\xi q_\infty) = \eta_\infty \sin q_\infty$.  
In the limit of small $\xi$ we find that $q_\infty$ will be close to $\pi$.  
Defining $\Delta := \pi - q$ we have the relation
\begin{equation}\label{eq:Delta}
   (1-4\xi^2) \sin\Delta_\infty
                            = (4-\xi^2) \sin[\xi(\pi-\Delta_\infty)] \ ,
\end{equation}
for $\Delta_\infty$.  This equation can be used to solve for $\Delta_\infty$
as a power series in $\xi$ with the result
\begin{equation}\label{eq:Deltaseries}
 \Delta_\infty = 4\pi\xi - 16\pi\xi^2 + \pi(79+10\pi^2)\xi^3 + \Oscr(\xi^4)\ . 
\end{equation}
An issue here is the radius of convergence of this series.  At this point we 
will content ourselves with the observation that for small enough $\xi$ the 
above series solution (\ref{eq:Deltaseries}) gives a good approximate solution 
to (\ref{eq:Delta}).  This is an indication that the radius of convergence is 
nonzero.  For general $\xi$ it turns out to be more effective to solve 
\eqref{eq:Delta} directly using a root finder routine.  To obtain the value of 
$\Delta$ at the causal limit, $\Delta_*(\xi)$, we proceed in an analogous way 
observing first that $\eta_* = (1-5\xi^2)/(5-\xi^2)$, where
$\eta_*$ is  defined by letting the sound speed at the center be equal to the
speed of  light, $u_{\rm c}=1$.  This leads to the equation
\begin{equation}
   (1-5\xi^2) \sin\Delta_* = (5-\xi^2) \sin[\xi(\pi-\Delta_*)] \ .
\end{equation}
Again we solve this equation numerically using a root finding routine. The 
functions $\Delta_\infty(\xi)$ and $\Delta_*(\xi)$ are plotted in figure 
\ref{fig:delta}.

%^^^^^^^^^^^^^^^^^^^^^^^^^^^^^^^^^^^^^^^^^^^^^^^^^^^^^^^^^^^^^^^^^^^^^^^^^^^^%
%----------------------------------------------------------------------------%

Although the parameter $q$ is a good variable to describe sequences of models 
with $\xi\neq0$ it is not a good variable in the limit $\xi \rightarrow 0$ 
since $q$ approaches the constant value $\pi$ in that limit.  We wish to 
obtain a variable which works well also in the $\xi\rightarrow 0$ limit.  
Since $\eta$ is a good variable for the $\xi=0$ models we look for a new 
variable $\sigma$ which has the property $\sigma \rightarrow \eta$ as $\xi 
\rightarrow 0$.  To that end we observe from \eqref{eq:q} that $\eta = \xi 
q/(\pi-q)$ when $\xi\ll1$.  This motivates the definition $\sigma := \xi q/ 
(\pi-q)$.  Then $\sigma \rightarrow \eta$ in the limit $\xi \rightarrow 0$ as 
required.  Also, $\sigma$ is a nonnegative increasing function of $q$ and 
hence is also an increasing function of $\eta$.  Finally we normalize $\sigma$ 
by defining $\hat\sigma := \sigma/(4\sigma_\infty)$ where $\sigma_\infty = \xi 
q_\infty/ (\pi-q_\infty) = \xi(\pi-\Delta_\infty)/ \Delta_\infty$.  Then 
$\hat\sigma \rightarrow \eta$ as $\xi \rightarrow0$ and 
$\hat\sigma_\infty=1/4$ for all $\xi$.  We have
\begin{equation}
   \hat\sigma = \frac{q\Delta_\infty}{4(\pi-q)(\pi-\Delta_\infty)} \ ,
\end{equation}
and solving for $q$ we obtain
\begin{equation}
   q = \displaystyle\frac{4\pi(\pi-\Delta_\infty)\hat\sigma}
            {\Delta_\infty + 4(\pi-\Delta_\infty)\hat\sigma} \ .
\end{equation}
Expressing the causal limit in terms of the normalized parameter gives 
\begin{equation}\label{eq:hatsigmaclimit}
   \hat\sigma_* = \displaystyle\frac{\Delta_\infty(\pi-\Delta_*)}
                                    {4\Delta_*(\pi-\Delta_\infty)} \ .
\end{equation}

%%%%%%%%%%%%%%%%%%%%%%%%%%%%%%%%%%%%%%%%%%%%%%%%%%%%%%%%%%%%%%%%%%%%%%%%%%%%%%
\subsection{Expressions for radius, mass and surface redshift} 
%^^^^^^^^^^^^^^^^^^^^^^^^^^^^^^^^^^^^^^^^^^^^^^^^^^^^^^^^^^^^^^^^^^^^^^^^^^^^%
\begin{figure}[!tbp]
\centering
\includegraphics{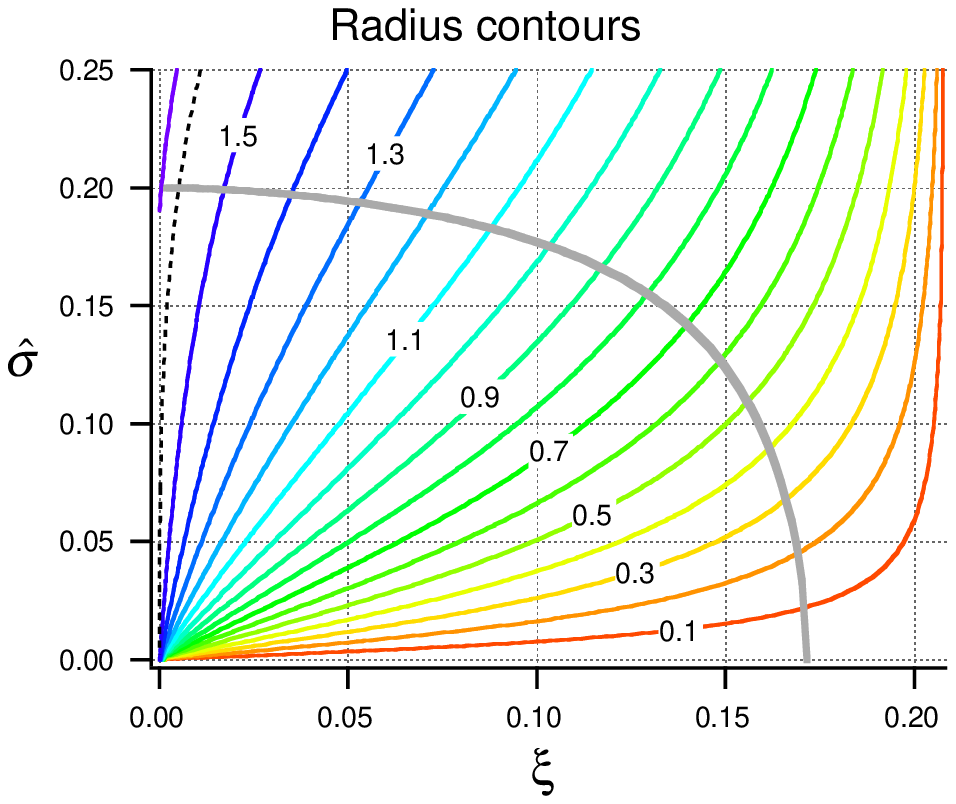}
\smallcaption{ Contours of stellar radii are shown for the \gbone\ family.  
The radii are indicated in units of the characteristic length scale $\ell$.  
The solid contours are equidistant.  The dotted contour has the value $r_{\rm 
s}/\ell = \pi/2 \approx 1.57$.  This is the minimum radius for Buchdahl's 
gaseous models.  The models with $\xi>0$ all have zero minimum radius 
reflecting their liquid-like equation of state.  The causal limit as defined 
by \eqref{eq:hatsigmaclimit} is indicated by the grey line.  Models below that 
line are everywhere causal, $u<c$.  }
\label{fig:radii}
\end{figure}
%----------------------------------------------------------------------------%
In the rest of the paper we set $R_\pm=0$.  This implies that the model is 
regular throughout the interior of the star.  To compute the radius we first 
use \eqref{eq:solution} to write
\begin{equation}\label{eq:r}
   r = W(R )= \begin{cases}
       \displaystyle\frac\ell2\cdot\frac{\sqrt{1-\xi^2}}{\sqrt{1-\eta^2}}
              \,\left[\xi^{-1}\sin(\omega_-R)+\eta\sin(\omega_+R)\right] \ ,
                      & \text{for $\xi\neq0$ \ ,}\\[15pt]
      \displaystyle\frac\ell2\cdot\frac1{\sqrt{1-\eta^2}}
                                       [\omega_+R + \eta\sin(\omega_+R)]            
                      & \text{for $\xi=0$ \ .}
        \end{cases}  
\end{equation}
where $\ell:= 1/\sqrt{\kappa a}$ is a characteristic length scale.  Evaluating 
this expression at the surface we find
\begin{equation}\label{eq:radius}
     r_{\rm s} = W_{\rm s}
      = \begin{cases}
           \displaystyle \frac{\ell(1+\xi)\sqrt{1-\xi^2}\sin q\sin(\xi q)}
                  {2\xi \sqrt{\sin^2q - \sin^2(\xi q)}}
                      & \text{for $\xi\neq0$ \ ,}\\[20pt]
           \displaystyle\frac{\pi\ell}{2\sqrt{1-\eta^2}}
                      & \text{for $\xi=0$ \ .}
        \end{cases}  
\end{equation}
The radius is smallest in the limit $\eta \rightarrow \xi$ (or equivalently $q 
\rightarrow 0$ if $\xi\neq0$).  It is evident from \eqref{eq:radius} that for 
Buchdahl's model the minimum radius is nonzero and is given by $(r_{\rm 
s})_{\rm min} = (\pi/2)\ell$.  The models with $\xi>0$ on the other hand all 
have zero minimum radius as expected from their liquid-like equation of state.  
A contour plot of radii for the \gbone\ family is shown in figure 
\ref{fig:radii}.

%^^^^^^^^^^^^^^^^^^^^^^^^^^^^^^^^^^^^^^^^^^^^^^^^^^^^^^^^^^^^^^^^^^^^^^^^^^^^%
\begin{figure}[!tbp]
\centering
\includegraphics{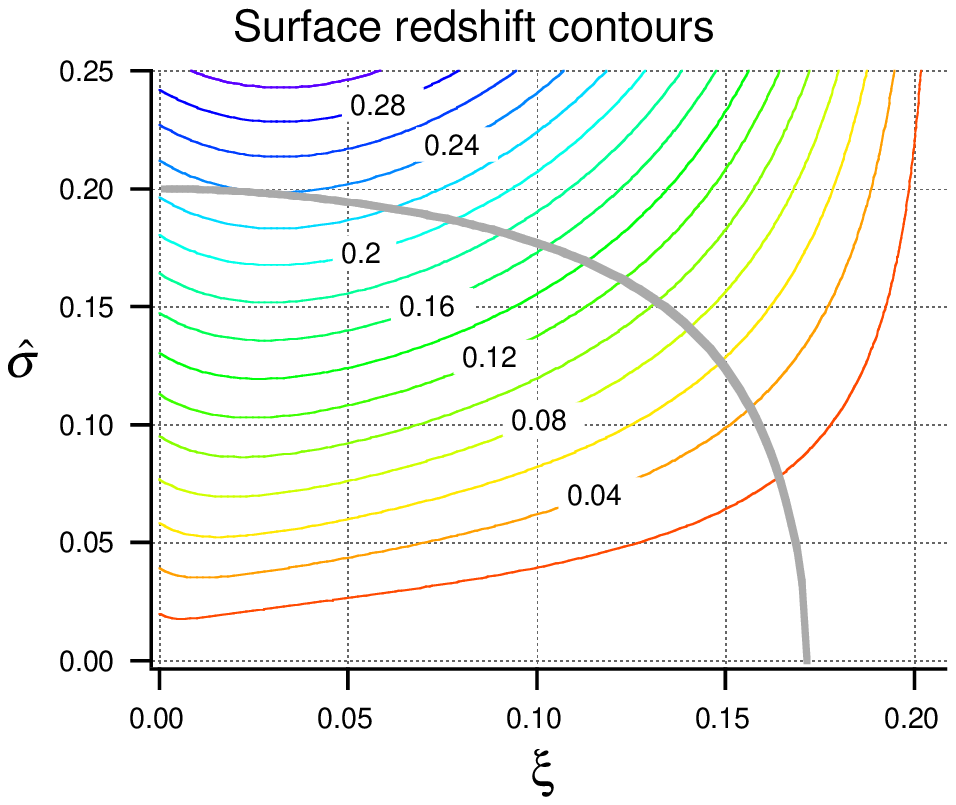}
\smallcaption{
Surface redshift contours are shown for all the \gbone\ models. The 
causal limit is indicated by the grey line. Models below that line are 
everywhere causal, $u<c$. }
\label{fig:redshift}
\end{figure}
\begin{figure}[!tbp]
\centering
\includegraphics{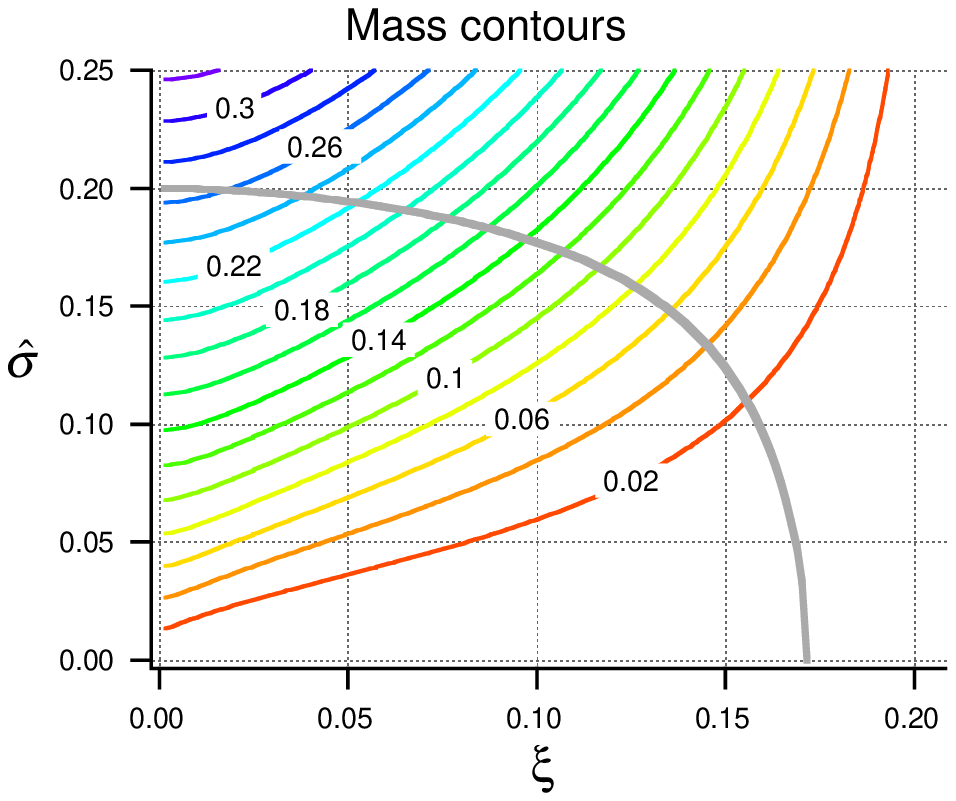}
\smallcaption{Mass contours are shown for all the \gbone\ models.  The masses 
are given in units of the characteristic length, $\ell$.  The causal limit is 
indicated by the grey line.  Models below that line are everywhere causal, 
$u<c$.  }
\label{fig:mass}
\end{figure}

%----------------------------------------------------------------------------%
To compute the redshift we begin by writing down an expression for $\beta$.  
For that we need
\begin{equation}
    W_{,R} = \begin{cases}
               (\cosh\zeta)\cos(\omega_-R) + (\sinh\zeta)\cos(\omega_+R)
                 & \text{for $\xi\neq0$ \ ,} \\[5pt]
                \cosh\zeta + (\sinh\zeta)\cos(\omega_+R)
                 & \text{for $\xi=0$ \ ,}
             \end{cases} 
\end{equation}
leading to
\begin{equation}
    (W_{,R})_{\rm s}
       = \begin{cases}
           (\cosh\zeta)\cos(\xi q) + (\sinh\zeta)\cos q
                = \sqrt{\displaystyle \frac{\sin[(1+\xi)q]}{\sin[(1-\xi)q]}}
                    & \text{for $\xi\neq0$\ ,} \\[10pt]
           e^{-\zeta} = \sqrt{\displaystyle\frac{1-\eta}{1+\eta}}
                    & \text{for $\xi = 0$ \ .}
         \end{cases}
\end{equation}
This gives
\begin{equation}
   \beta = \begin{cases}
             \case12\left\{1- \displaystyle\frac
                  {(1-\xi)\sin[(1+\xi)q]}{(1+\xi)\sin[(1-\xi)q]} \right\}
               & \text{for $\xi\neq0$} \ ,\\[15pt]
             \displaystyle\frac{\eta}{1+\eta} & \text{for $\xi=0$} \ .  
   \end{cases}
\end{equation}
The expression for the redshift becomes
\begin{equation}\label{eq:redshift}
   1+z_{\rm s} = \begin{cases}
                  \sqrt{\displaystyle\frac
                  {(1+\xi)\sin[(1-\xi)q]}{(1-\xi)\sin[(1+\xi)q]}}
                                         & \text{for $\xi\neq0$} \ ,\\[20pt]
          \sqrt{\displaystyle\frac{1+\eta}{1-\eta}} & \text{for $\xi=0$} \ .  
   \end{cases}
\end{equation}
Comparison with the Doppler relation between redshift and velocity shows that 
for Buchdahl's model $\eta$ is exactly equal to the corresponding velocity.  
The maximum equivalent velocity for Buchdahl's model is therefore $v = \eta = 
1/4$ corresponding to a maximum redshift of $z_{\rm s} = \sqrt{5/3}-1 \approx 
0.291$.  Restriction to causal models ($v = \eta<1/5$) gives a maximum 
redshift $z_{\rm s} = \sqrt{3/2}-1 \approx 0.225$.  A contour plot of surface 
redshifts for the \gbone\ family is shown in figure \ref{fig:redshift}.  The 
maximum value of the surface redshift can be computed numerically to be 
$z\approx 0.31$ which occurs for $\xi\approx 0.032$.  The corresponding 
compactness parameter is $\alpha=1/\beta \approx 4.8$.  Therefore there are no 
ultracompact models in the \gbone\ family.  Turning now to the mass we find 
that it can be written as
\begin{equation}
     M = \begin{cases}
           \displaystyle\frac{\ell(1+\xi)\sqrt{1-\xi^2}\sin q\sin(\xi q)}
                  {4\xi \sqrt{\sin^2q - \sin^2(\xi q)}} \left\{1-
           \displaystyle\frac{(1-\xi)\sin[(1+\xi)q]}
                       {(1+\xi)\sin[(1-\xi)q]} \right\}
                           & \text{for $\xi\neq0$ \ ,} \\[20pt]
           \displaystyle\frac{\pi\eta\ell}{2(1+\eta)\sqrt{1-\eta^2}}
                           & \text{for $\xi=0$ \ .}
         \end{cases}
\end{equation}
A contour plot of masses for the \gbone\ family is shown in figure 
\ref{fig:mass}.
%^^^^^^^^^^^^^^^^^^^^^^^^^^^^^^^^^^^^^^^^^^^^^^^^^^^^^^^^^^^^^^^^^^^^^^^^^^^^%
%----------------------------------------------------------------------------%

The central values of the pressure and energy density are given by
\begin{equation}
     p_{\rm c} = \frac{4a(\eta^2-\xi^2)}{(1-\xi^2)(1+\eta)^2} \ ,\qquad
     \rho_{\rm c} = \frac{4a[\eta(2-3\eta)+\xi^2(3-2\eta)]}
                         {(1-\xi^2) (1+\eta)^2} \ .
\end{equation}
The central values of the adiabatic index and sound speed are
\begin{equation}
   \gamma_{\rm c} = \frac{2(1-\eta)(\eta+\xi^2)^2}
                         {(\eta^2-\xi^2)[1-4\eta-(4-\eta)\xi^2]} \ ,\qquad
   u_{\rm c} = \frac{\eta+\xi^2}
                                  {1-4\eta-(4-\eta)\xi^2} \ .
\end{equation}
The surface density is given by
\begin{equation}
   \rho_{\rm s} = \frac{8a\xi}{(1+\xi)^2} \ .
\end{equation}

%%%%%%%%%%%%%%%%%%%%%%%%%%%%%%%%%%%%%%%%%%%%%%%%%%%%%%%%%%%%%%%%%%%%%%%%%%%%%%
\subsection{Pressure and density profiles}
\begin{figure}[!tbp]
 \centering
  \subfigure[Density profiles.]
            {\label{fig:densB}
              \psfrag{rhoB_bottom}[][]{\Large $r/\ell$}
              \psfrag{rhoB_left}[][]{\Large $\rho/\rho_0$}
             \includegraphics[width=0.45\textwidth]{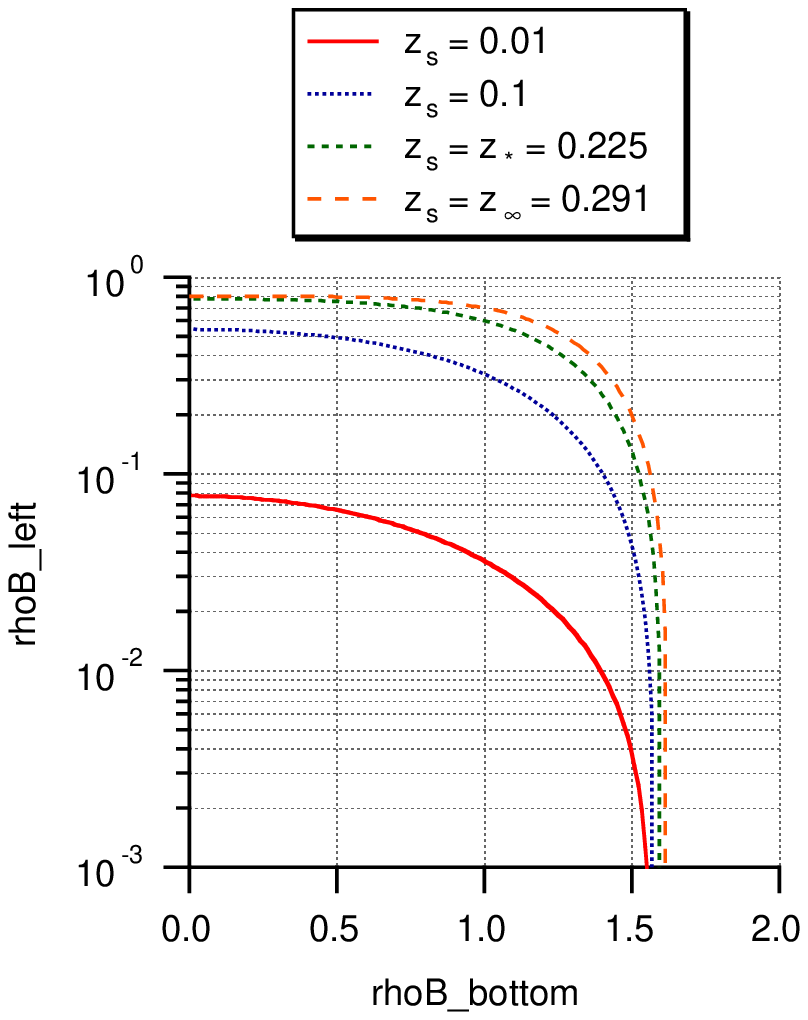}}
  \subfigure[Pressure profiles.]
            {\label{fig:pB}
              \psfrag{pB_bottom}[][]{\Large $r/\ell$}
              \psfrag{pB_left}[][]{\Large $p/p_0$}
             \includegraphics[width=0.45\textwidth]{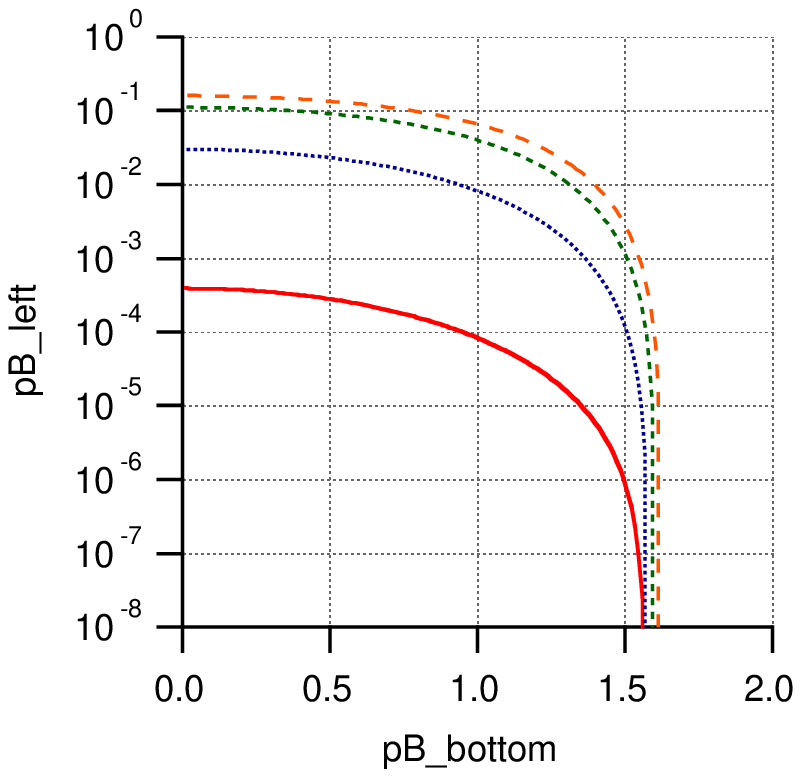}}
  \smallcaption{Density and pressure profiles for Buchdahl's model ($\xi=0$).  
  The density and pressure are given in units of $\rho_0= 
  \kappa^{-1}\ell^{-2}$ and $p_0= \kappa^{-1}\ell^{-2}$.  The profiles are 
  given for four representative values of the surface redshift ranging from 
  the mariginally relativistic, $z=0.01$, to the most extreme relativistic 
  case, $z=z_\infty = \sqrt{5/3}-1 \approx 0.291$.  The radius takes values in 
  the very narrow range, $\pi/2 \leq r/\ell <2\pi/\sqrt{15}$, or approximately 
  $1.571 \leq r/\ell < 1.622$.  }
 \label{fig:profilesB}
\end{figure}

\begin{figure}[!tbp]
 \centering
  \subfigure[Density profiles.]
            {\label{fig:densII}
             \includegraphics[width=0.45\textwidth]{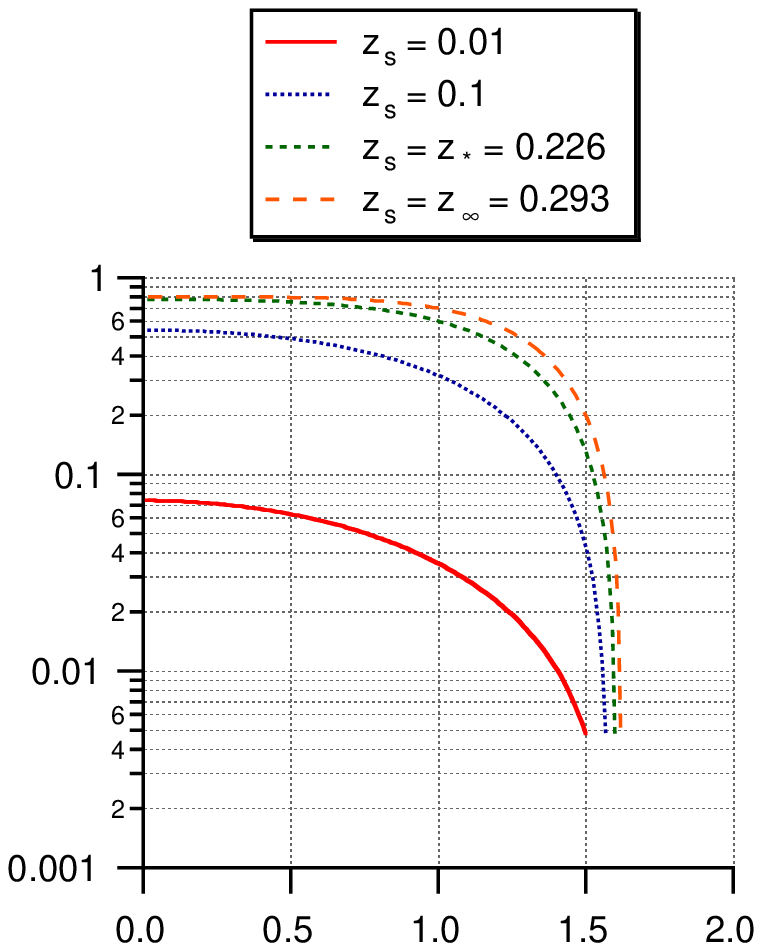}}
  \subfigure[Pressure profiles.]
            {\label{fig:pII}
             \includegraphics[width=0.45\textwidth]{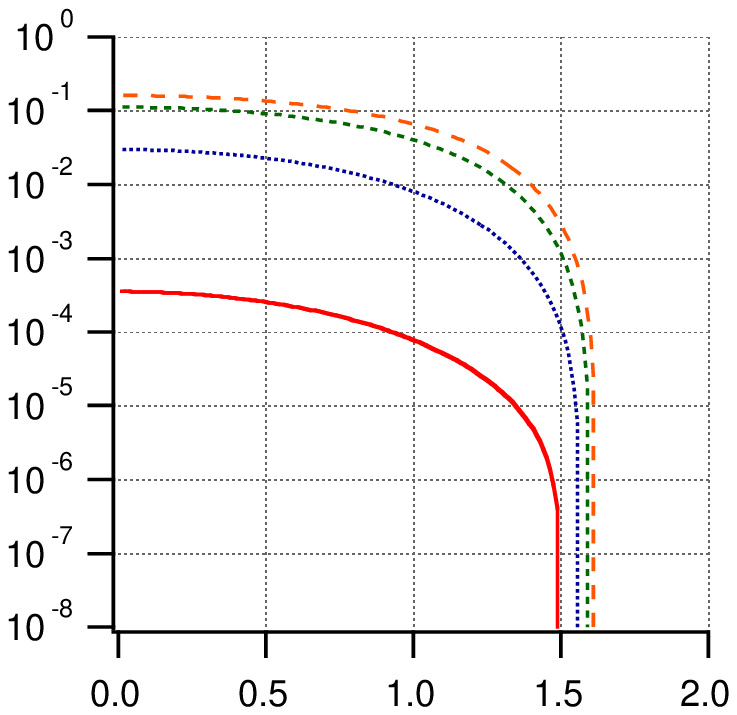}}
  \smallcaption{Profiles for the type II \gbone\ models ($\xi=0.0006$).  Axes 
  and units are as in figure \ref{fig:profilesB}.  The profiles are similar to 
  those of the Buchdahl model except that the density is nonzero at the 
  surface. }
 \label{fig:profilesII}
\end{figure}

%%%%%%%%%%%%%%%%%%%%%%%%%%%%%%%%%%%%%%%%%%%%%%%%%%%%%%%%%%%%%%%%%%%%%%%%%%%%%%
The pressure and density functions are given in \eqref{eq:eqstate} as
functions of $Z$. Expressing this function in terms of the normalized radial
variable $\hat R := R/R_{\rm s}$ we find
\begin{equation}
   Z(\hat R) = \frac{1-\phi(\hat R)}{1+\phi(\hat R)} \ ,
\end{equation}
where
\begin{equation}
   \phi(\hat R) := X/T = \begin{cases}
                  \displaystyle\frac{\xi\sin(\xi q)\sin(q\hat R)}
                        {\sin q \sin(\xi q\hat R)} \ .
                      & \text{for $\xi\neq0$ \ ,}\\[15pt]
                  \displaystyle\frac{\eta\sin(\pi\hat R)}{\pi\hat R}
                      & \text{for $\xi=0$ \ .}
               \end{cases}
\end{equation}
These formulas together with \eqref{eq:eqstate} give $p$ and $\rho$ as 
functions of $\hat R$.  We wish to plot the pressure and density versus 
Schwarzschild's radial variable which was given in equation \eqref{eq:r}.  The 
result for Buchdahl's model is shown in figure \ref{fig:profilesB}.  A 
striking feature is the small variation in the radius.  In going from a low 
mass Newtonian model with $z=0$ up to the most compact relativistic model with 
redshift $z=0.291$ the variation in the radius is only about 3\%.  The 
profiles for a representative of the type II GB1 family are shown in figure 
\ref{fig:profilesII}.

%Unlike the profiles calculated by Herold and Neugebauer \cite{hn:varstar} the 
%type II GB1 profiles do not overlap for different values of the surface 
%redshift.  Since the equations of state are very similar in the outer regions 
%of the star, this difference must depend on the equation of state closer to 
%the core.

%^^^^^^^^^^^^^^^^^^^^^^^^^^^^^^^^^^^^^^^^^^^^^^^^^^^^^^^^^^^^^^^^^^^^^^^^^^^^%
%----------------------------------------------------------------------------%
%^^^^^^^^^^^^^^^^^^^^^^^^^^^^^^^^^^^^^^^^^^^^^^^^^^^^^^^^^^^^^^^^^^^^^^^^^^^^%
%----------------------------------------------------------------------------%
%^^^^^^^^^^^^^^^^^^^^^^^^^^^^^^^^^^^^^^^^^^^^^^^^^^^^^^^^^^^^^^^^^^^^^^^^^^^^%
\begin{figure}[!tbp]
 \centering
  \subfigure[Density profiles.]
            {\label{fig:densIII}
             \includegraphics[width=0.45\textwidth]{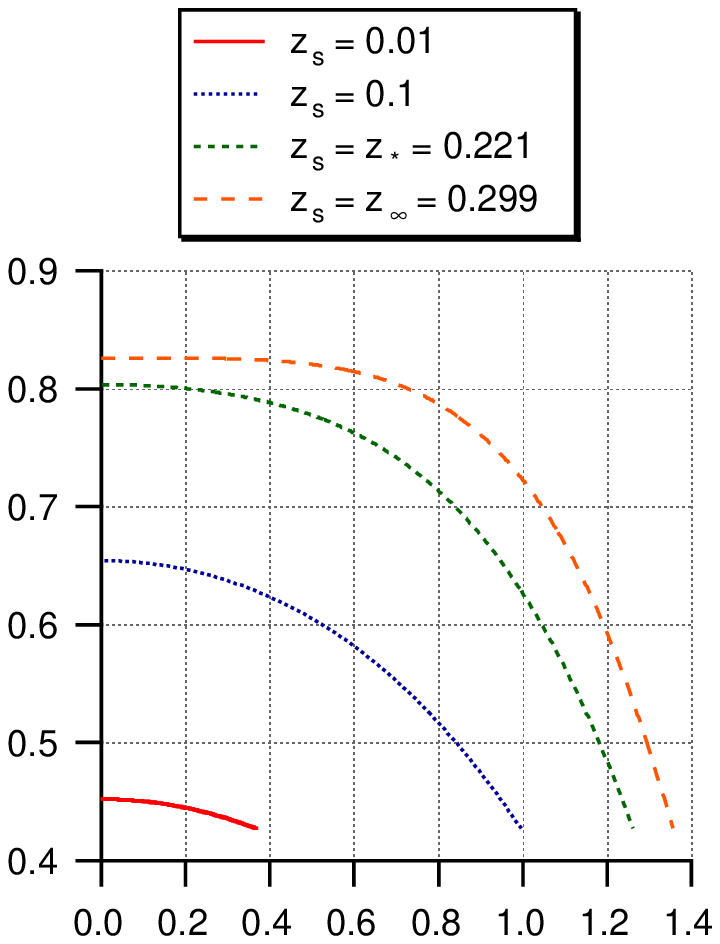}}
  \subfigure[Pressure profiles.]
            {\label{fig:pIII}
             \includegraphics[width=0.45\textwidth]{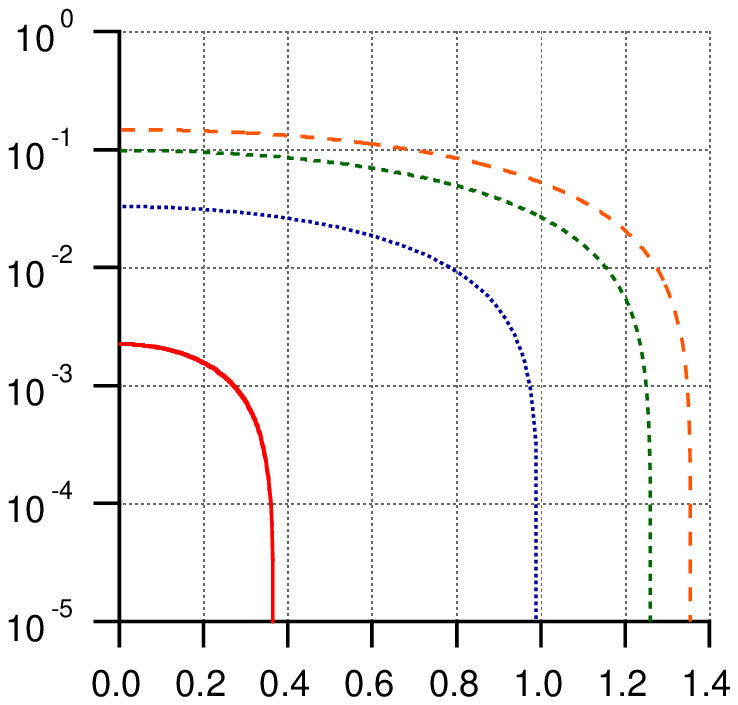}}
  \smallcaption{Profiles for type III \gbone\ models ($\xi=0.06$). Axes 
  and units are as in figure \ref{fig:profilesB}.}
 \label{fig:profilesIII}
\end{figure}
%----------------------------------------------------------------------------%
%^^^^^^^^^^^^^^^^^^^^^^^^^^^^^^^^^^^^^^^^^^^^^^^^^^^^^^^^^^^^^^^^^^^^^^^^^^^^%
\begin{figure}[!tbp]
 \centering
  \subfigure[Density profiles.]
            {\label{fig:densIV}
             \includegraphics[width=0.45\textwidth]{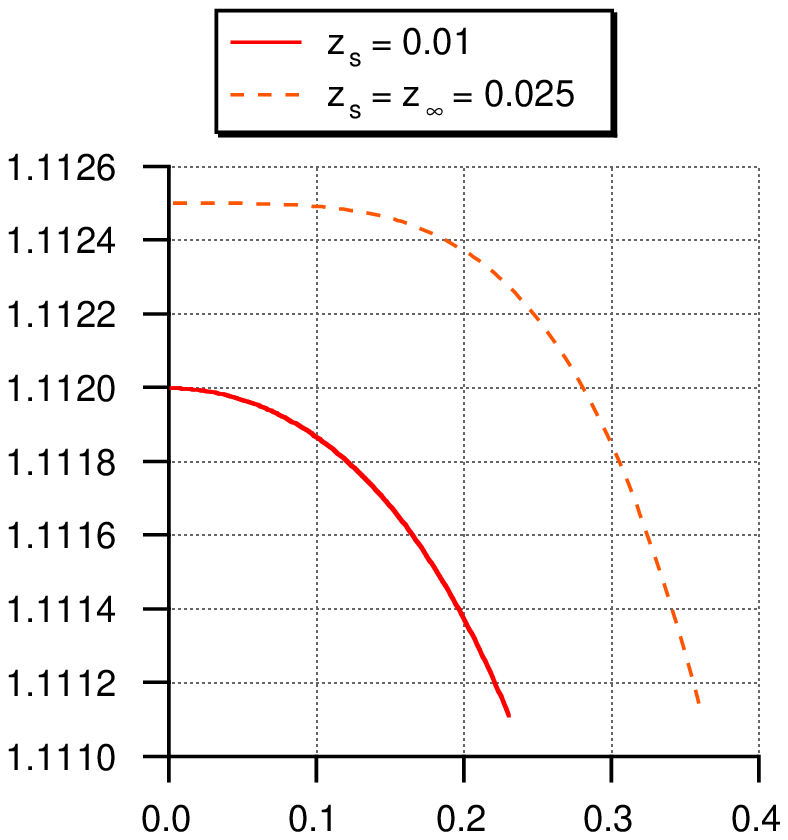}}
  \subfigure[Pressure profiles.]
            {\label{fig:pIV}
             \includegraphics[width=0.45\textwidth]{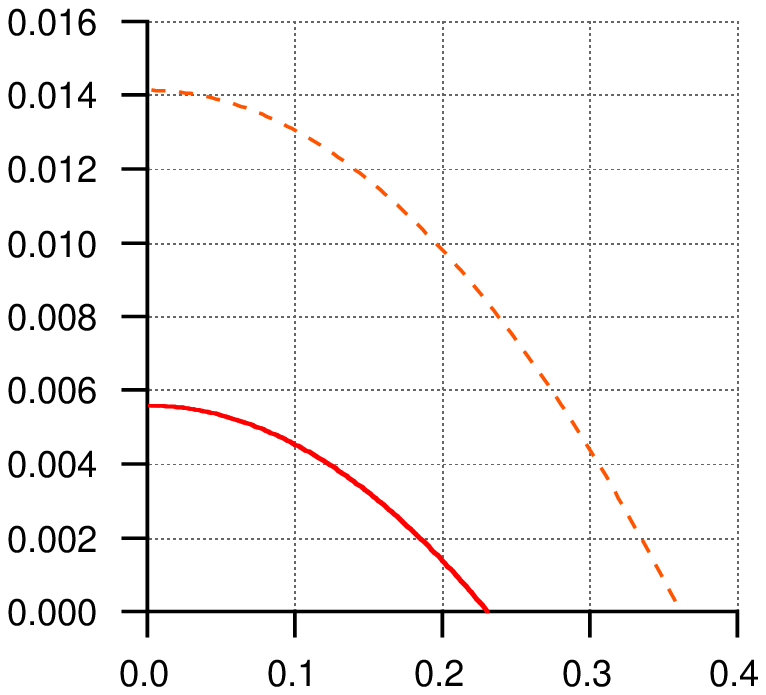}}
  \smallcaption{Profiles for type IV \gbone\ models ($\xi=0.2$). Axes 
  and units are as in figure \ref{fig:profilesB}.}
 \label{fig:profilesIV}
\end{figure}
%----------------------------------------------------------------------------%

%%%%%%%%%%%%%%%%%%%%%%%%%%%%%%%%%%%%%%%%%%%%%%%%%%%%%%%%%%%%%%%%%%%%%%%%%%%%%%
\section{Concluding remarks}
We have described the 2-parameter \gbone\ family of exact spherically 
symmetric relativistic stellar models with a physically reasonable equation of 
state.  One of the parameters represents the scaling invariance while the 
second parameter can be interpreted as a measure of the stiffness.  By 
restricting the range of the stiffness parameter, one obtains a subfamily for 
which the sound speed is subluminal throughout the stellar interior.  By 
varying the equation of state parameters it is possible to obtain a quite 
accurate fitting to the \hw\ equation of state corresponding to the outer 
region of a neutron star.

Being a realistic model of a neutron star near the surface the \gbone\ family 
is a good candidate for a static limit of a rotating neutron star model.  The 
main difficulty in trying to spin up a static perfect fluid model is to 
preserve the perfect fluid property.  Technically this means that the Ricci 
tensor should have the correct algebraic type (Segr\'e characteristic 
A1[(111),1)], see reference \cite{ksmh:exact}) and in addition contain only a 
single arbitrary function.  A recent result \cite{ds:nj} seems to indicate 
that the Newman-Janis algorithm would always destroy the perfect fluid 
property.  This could mean that a rotating perfect fluid model would not have 
a Kerr exterior.

The \gbone\ family is the only exact general physical \sss\ solution which is 
expressible in terms of elementary functions for all values of the 
integrations constants.  For Schwarzschild's interior solution, for example, 
only the regular models can be described in terms of elementary functions 
while the singular cases involve elliptic functions.  It should be noted that 
the singular models have considerable physical interest.  First they can be 
used in the regions outside the core of the star.  Also, they are needed for 
obtaining a better understanding of the space of solutions as well as of the 
possibilities for global behaviour.

%The latter point has gotten a renewed interest in connection with suggestions 
%for black holes with short hair (see {\em e.g. ??}).

%conical singularity

%%%%%%%%%%%%%%%%%%%%%%%%%%%%%%%%%%%%%%%%%%%%%%%%%%%%%%%%%%%%%%%%%%%%%%%%%%%%%%
%%%%%%%%%%%%%%%%%%%%%%%%%%%%%%%%%%%%%%%%%%%%%%%%%%%%%%%%%%%%%%%%%%%%%%%%%%%%%%
\appendix
\section*{Appendix}

%%%%%%%%%%%%%%%%%%%%%%%%%%%%%%%%%%%%%%%%%%%%%%%%%%%%%%%%%%%%%%%%%%%%%%%%%%%%%%
\section* {Useful formulas}
\label{app:form}

First some definitions:
\begin{equation}
     \omega_\pm := \sqrt{2(\delta\pm1)\kappa a} \ ,\qquad
     \ell := 1/\sqrt{\kappa a} \ ,\qquad
     \xi := \omega_-/\omega_+ = \sqrt{\frac{\delta-1}{\delta+1}} \ .
\end{equation}
We then have the following identities
\begin{equation}
     \delta = \frac{1+\xi^2}{1-\xi^2} \ ,\qquad
     \delta+1 = \frac2{1-\xi^2} \ ,\qquad
     \delta-1 = \frac{2\xi^2}{1-\xi^2} \ ,\qquad
     \delta-\sqrt{\delta^2-1} = \frac{1-\xi}{1+\xi} \ ,
\end{equation}
and
\begin{equation}
     \omega_+{}^{-1} = \case12\ell\sqrt{1-\xi^2} \ ,\qquad
     \omega_-{}^{-1} = \case12\ell\xi^{-1}\sqrt{1-\xi^2} \ .
\end{equation}
\begin{equation}
     \eta := \tanh\zeta \ ,\qquad
     e^{2\zeta} = \frac{1+\eta}{1-\eta} \ ,\qquad
     \cosh\zeta = \frac1{\sqrt{1-\eta^2}} \ ,\qquad
     \sinh\zeta = \frac\eta{\sqrt{1-\eta^2}} \ .
\end{equation}

\clearpage

\bibliographystyle{prsty}
%\bibliography{kr}

\end{document}